\documentstyle[12pt]{article}
\begin{document}
\def\omt{\tilde{\omega}}
\def\ti{\tilde}
\def\o{\Omega}
\def\bchi{\bar\chi^i}
\def\In{{\rm Int}}
\def\ba{\bar a} 
\def\w{\wedge}
\def\ep{\epsilon}
\def\k{\kappa}
\def\Tr{{\rm Tr}}
\def\ST{{\rm STr}}
\def\ss{\subset}
\def\rn{\vert \alpha\vert^2}
\def\bi{\bibitem}
\def\ot{\oti\def\om{\omega}
\dmes}
\def\bc{{\bf C}}
\def\bz{{\bf Z}}
\def\ptp{\stackrel{\otimes}{,}}
\def\br{{\bf R}}
\def\de{\delta}
\def\od{\sqrt{2}}
 \def\bt{\beta}
 \def\ve{\vert}
\def\al{\alpha}
 \def\hal{\hat{\alpha}}
 \def\nn{\nonumber}
\def\la{\langle}
\def\ra{\rangle}
\def\Ga{\Gamma}
 \def\ul{\underline}
\def\st{\stackrel{\wedge}{,}}
\def\stv{\stackrel{\wedge}{\vert}}
\def\th{\theta}
\def\lm{\ti\lambda}
\def\U{\varsigma}
\def\jp{{1\over 2}}
\def\js{{1\over 4}}
\def\mb{\ _b\mu}
\def\d{\partial}
\def\tr{\triangleright}
\def\trl{\triangleleft}
\def\pl{\Pi_B^L}
\def\pr{\Pi_B^R}
\def\pa{\Pi_B^a}
\def\tpa{\ti\Pi_B^a}
\def\d{\partial}
\def\bq{\}_{P}}
\def\be{\begin{equation}}
\def\ee{\end{equation}}
\def\bea{\begin{eqnarray}}
\def\eea{\end{eqnarray}}
\def\D{{\cal D}}
\def\A{{\cal A}}
\def\G{{\cal G}}
\def\K{{\cal K}}
\def\H{{\cal H}}
\def\L{{\cal L}}
\def\P{{\cal P}}
\def\N{{\cal N}}
\def\R{{\cal R}}
\def\B{{\cal B}}
\def\T{{\cal T}}
\def\bT{\bar{\cal T}}
\def\rbo{\rightline{$\Box$}}
\def\F{{\cal F}}
\def\n{{1\over n}}
\def\si{\sigma}
\def\ta{\tau}
\def\ov{\over}
\def\lm{\lambda}
\def\Lm{\Lambda}
\def\lpb{{\bf \{}}
\def\rpb{{\bf \}}}

\def\pih{\hat{\pi}}
\def\noi{\noindent}

\def\U{\varsigma}
\def\e{\varepsilon}
\def\bt{\beta}
\def\ga{\gamma}
\def\om{\omega}
\def\Om{\Omega}

\def\D{{\cal D}}
\def\C{{\cal C}}
\def\G{{\cal G}}
\def\H{{\cal H}}
\def\R{{\cal R}}
\def\B{{\cal B}}
\def\K{{\cal K}}
\def\T{{\cal T}}
\def\M{{\cal M}}
\def\bT{\bar{\cal T}}
\def\F{{\cal F}}
\def\n{{1\over n}}
\def\si{\sigma}
\def\ta{\tau}
\def\ot{\otimes}
  \def\lpk{L_{pol}K}
\def\lpkc{L_{pol}K^{\bf C}}
\def\ve{\vert}
\def\nr{\nabla^R}
\def\nl{\nabla^L}
\def\pih{\hat{\pi}}

\def\e{\varepsilon}
\def\bt{\beta}
\def\ga{\gamma}

\begin{titlepage}

 \begin{flushright}
{}~
IML 2006-22\\
math-ph/0611066
\end{flushright}

\vspace{3cm}
\begin{center}
{\large \bf  $q\to \infty$ limit of the quasitriangular WZW model}\\ 
[50pt]{\small
{\bf C. Klim\v{c}\'{\i}k }
\\ ~~\\Institute de math\'ematiques de Luminy,
 \\163, Avenue de Luminy, 13288 Marseille, France}

\vspace{0.5cm}

\begin{abstract}
\noindent
We study the $q\to\infty$ limit of the $q$-deformation of the WZW model on a compact simple and simply
connected  target  Lie group.  We show that 
 the commutation relations of the $q\to\infty$ current algebra are underlied by certain 
 affine Poisson structure on the group of holomorphic maps from the disc into the complexification of the target group.  The Lie algebroid corresponding to this affine Poisson
 structure can be integrated to a global symplectic groupoid which  turns out  to be
  nothing but the phase space of the $q\to\infty$ limit of the $q$-WZW model.
 We    also  show that this symplectic grupoid  admits a
chiral decomposition compatible with its  (anomalous) Poisson-Lie symmetries.  Finally,  we dualize the chiral theory in a remarkable way
and we evaluate the exchange   relations for the $q\to\infty$  chiral WZW fields  
in both the original and the dual pictures. \end{abstract}

\end{center}
\end{titlepage}

\section{Introduction}

The goal of the present paper is to study the $q\to \infty$ limit of the quasitriangular WZW
model \cite{K04}, which  is the $q$-deformation of the standard 
WZW model \cite{W}  (the standard WZW model  corresponds to the limit $q\to 1$).  It will turn out  that  in the $q\to\infty$
limit the quasitriangular WZW model simplifies considerably, while enjoying   chiral decomposability and  other agreeable features
of its finite $q$ analogues.   In particular, we shall find  that  the elliptic $r$-matrices, which characterize the chiral exchange relations,  become trigonometric in the $q\to\infty$ limit.
On the other hand, we shall show, rather remarkably, that inspite of this simplification, the  symmetry pattern of the $q\to\infty$ model is richer and more intricate than in the case of finite $q$. 

\vskip1pc

\noindent  
 An important tool  for analyzing the rich structure of the quasitriangular WZW model
is  the theory of affine Poisson groups and the associated concept  of 
anomalous  Poisson-Lie symmetry of dynamical systems.  Affine Poisson groups  have been introduced by Dazord and Sondaz \cite{DS} as generalizations
of Poisson-Lie groups.  To every affine Poisson bivector  $\Pi^*$ on a Lie group $G^*$ there are
associated two Poisson-Lie bivectors $\Pi^*_L $, $\Pi^*_R$ on $G^*$. We shall denote by $G_L$ and $G_R$ the dual Poisson-Lie groups of the Poisson-Lie groups $(G^*,\Pi^*_L)$ and $(G^*,\Pi^*_R)$, respectively. 
 In \cite{Lu1,Lu2}, Lu has pointed out the following fact: if $\mu$ is a smooth Poisson map
 from a symplectic manifold $P$ into $(G^*,\Pi^*)$ then there exists a pair of (infinitesimal) Poisson-Lie
 symmetries of $P$ with the symmetry groups $G_L$ and $G_R$.   
   If $\Pi^*$ is itself a Poisson-Lie
 bivector, then  $\Pi^*=\Pi^*_L=\Pi^*_R$  and $G_L=G_R\equiv G$. The  Poisson map 
 $\mu:P\to (G^*,\Pi^*)$ is then said to be an {\it equivariant} moment map of the pair of the $G$ 
 Poisson-Lie symmetries  of $P$. Note that we still speak about the {\it pair} of symmetries
 because   even in the equivariant case the action of
 the symmetry group  $G_L=G$ on $P$ need not coincide with the action of $G_R=G$ on $P$.
 If $\Pi^*$ is Poisson but not a Poisson-Lie structure, then we say that the Poisson map 
 $\mu:P\to (G^*,\Pi^*)$ is an  {\it anomalous}   moment map generating  the pair of anomalous
 $G_L$ and $G_R$ Poisson-Lie symmetries  of $P$.  

 \vskip1pc

 \noindent  We remark that  even  in the anomalous case,
  $G_L$ and $G_R$  may be isomorphic as 
  the Lie groups. This happens for 
\noindent the quasitriangular WZW model  if  
$q$ is finite.   (In the limit $q\to 1$, the symmetry structure of the model simplifies
 even more: not only $G_L$ is isomorphic to $G_R$ but also the $G_L=G$ action on the phase space $P$ of the model 
 coincides with the $G_R=G$ action.)   We shall show in this article that, at the opposite
 end of the range of the deformation parameter $q$,  the full structural richness of the concept of the anomalous Poisson-Lie symmetry is realized. We mean by this that in the limit $q\to\infty$ the group $G_L$ is   no longer isomorphic to the group $G_R$.  This fact leads  to an interesting duality  
 in  the $q\to\infty$ WZW theory which  interchanges the roles of the symmetry groups
 $G_L$ and $G_R$ in the description of the dynamics of the model.

 \vskip1pc

\noindent  The plan  of the paper is as follows. In Sections 2 and 3, we review  the elements of the
theory of  the affine Poisson groups and of  the anomalous Poisson-Lie symmetries.
In particular, we shall identify the moment maps of the anomalous Poisson-Lie symmetries
of the symplectic grupoids integrating  the affine Poisson groups. In Section 4, we turn our attention
to the Poisson-Lie anomalies of loop group symmetries.  The  phase space of the $q\to\infty$ 
limit of the
quasitriangular WZW model \cite{K04} will be shown to have the structure of the symplectic groupoid   integrating
certain affine Poisson structure on the  group  $G^*$ whose elements are holomorphic maps from the disc into the complexification of the target group of the WZW model. We shall also establish
the chiral decomposition of the  $q\to\infty$ WZW theory and show
that the chiral model is also Poisson-Lie symmetric.  Moreover, 
we shall work out in detail the exchange (braiding)
relations and $q\to \infty$ current algebra relations in  the 
  chiral sector. On the top of it, we shall describe also the remarkable duality of the chiral $q\to\infty$ WZW model
  which permits to express the
  symplectic structure of the model in terms of the two dual groups $G_L$ and $G_R$ naturally associated to the affine Poisson structure on $G^*$.

 \vskip 1pc

 \section{Affine Poisson groups}

 \noi The   affine Poisson groups were introduced by Dazord 
and Sondaz \cite{DS}  and  the basic facts about them
can be found in \cite{DS,Lu1,Za}. We give here a short summary of some of the results contained
in those papers. 
Note that all manifolds, maps or sections of bundles will be understood to be  smooth,
 moreover, slightly abusing terminology,  we shall often speak about  e.g. vectors and forms
 instead of vector fields and form fields, respectively.

 \medskip

 \noi A manifold $P$ equipped with a   bivector $\Pi$    is called Poisson if the following bracket
 defines a Lie algebra commutator on   the space of   functions on $P$  
 $$ \{x,y\}=\Pi(dx,dy).\nn$$
 Here $x,y$ are any  functions on $P$. Note that  a map $\mu$ from another Poisson manifold $P'$
 into $P$ is called  Poisson if it 
   intertwines the corresponding Poisson brackets, i.e. $$ \mu^*\{x,y\}=\{\mu^*x,\mu^*y\}'.\nn$$
     It is 
  well-known \cite{Do,MM,Ko,DS} that the Poisson bivector $\Pi$
 induces a Lie algebra structure also on the space   consisting of  $1$-form fields  on $P$. 
 It is  defined by the following formula \cite{DS}:
 $$\{\al_1,\al_2\}=\L_{\Pi(\al_1,.)}\al_2 - \L_{\Pi(\al_2,.)}\al_1-d(\Pi(\al_1,\al_2)), \eqno(2.1)
$$ 
 where $\L_v\al$ denotes the Lie derivative of the $1$-form $\al$ with respect to the 
vector field $v$.

\medskip

\noindent {\bf Definition 1}
\noi Let $R(G^*)$    be   the space of right-invariant  
$1$-forms
on   a group
manifold $G^*$.  An affine Poisson structure  on $G^*$ is 
a Poisson bivector $\Pi^*$ such that  the bracket (2.1) defines the Lie algebra structure
on $R(G^*)$.   When the affine Poisson bivector vanishes at the  
unit element $e^*\in G^*$, the group $(G^*,\Pi^*)$ is called the Poisson-Lie group.

\rightline{\#}

\noi  \noindent{\bf Lemma 1} A Poisson bivector $\Pi^*$ on  $G^*$ defines an affine Poisson structure
if and only if  the bracket (2.1) defines a Lie algebra structure
on the space $L(G^*)$ of left-invariant  
$1$-forms on   $G^*$.    

  \rightline{\#}

 \noi   We remark that both spaces $R(G^*)$ and $L(G^*)$   can be naturally 
  identified with the dual vector space $\G$ of the Lie algebra $\G^*$ of $G^*$.
  Thus the affine Poisson structure on $G^*$ defines two (not necessarily) isomorphic
  Lie algebra structures on $\G$, we denote them $\G_L=(\G,[.,.]_L)$ and $\G_R=(\G,[.,.]_R)$.

\medskip

  \noi\noindent{\bf Lemma 2} Let $(G^*,\Pi^*)$ be an affine Poisson group and denote by $M$
  the value of the bivector $\Pi^*$ at the group unit $e^*$.  Then bivectors $$
  \Pi_L^*\equiv \Pi^*-L_*M, \quad 
  \Pi_R^*\equiv \Pi^*-R_*M\nn$$are both  Poisson-Lie bivectors on $G^*$ and
  a bivector $$ \Pi^*_{op}\equiv \Pi^*-R_*M-L_*M\nn$$ is an affine Poisson bivector on $G^*$.
  
  \rightline{\#}

   \noi   The symbols $L_*$ and $R_*$ stand for the right and left translations from
   $e^*$ onto the whole group manifold $G^*$. The bivectors $\Pi^*_L$ and $\Pi^*_R$
   are referred to as the left and right  Poisson-Lie structures associated to the affine
   Poisson structure $\Pi^*$; $\Pi^*_{op}$ is called the opposed affine Poisson structure with 
   respect to $\Pi^*$.  It turns out that $(\Pi^*_{op})_{op}= \Pi^*$ and
   $\Pi^*_L$ ($\Pi^*_R$) is the right(left) associated Poisson-Lie structures to the affine
   Poisson structure $\Pi^*_{op}$.

 \medskip

\noindent{\bf Lemma 3} To every 
  affine Poisson group   $(G^*,\Pi^*)$ it can be associated a  Lie algebra $\D$
  such that:

\noi  1) There exist three injective Lie algebra homomorphisms $\varsigma:\G^*\to\D$,
$\varsigma_L:\G_L\to \D$ and $\varsigma_R:\G_R\to D$ such that $\D=\varsigma(\G^*)\stackrel{.}{+}\varsigma_L(\G_L)$ and also $\D=\varsigma(\G^*)\stackrel{.}{+}\varsigma_R(\G_R)$
.

 \noi 2) There is an $Ad$-invariant non-degenerate symmetric bilinear form $(.,.)_\D$ on $D$ which vanishes when restricted 
 to each of the subalgebras  $\varsigma(\G^*)$,$\varsigma_L(\G_L)$ and $\varsigma_R(\G_R)$.

\rightline{\#}
\noi Note that the symbol $\stackrel{.}{+}$ means the direct sum of vector spaces but not
necessarily the direct sum of Lie algebras (the latter denoted usually by $\oplus$).

\medskip

   \noi   The affine Poisson structure on a simply connected group $G^*$ can be
 completely reconstructed from  the Lie algebra $\D$  by considering 
 a simply connected group $D$ whose Lie algebra is $\D$.  The explicite formula  for $\Pi^*$ is as follows
 $$R_{{g^*}^{-1}}\Pi^*(g^*)(\xi_1,\xi_2)= -(Ad_{{ \varsigma(g^*)}^{-1}}\varsigma_L(\xi_1), p_R  Ad_{{\varsigma (g^*)}^{-1}}\varsigma_L(\xi_2))_\D, \  g^*\in G^*,   \xi_1,\xi_2\in \G,\eqno(2.2)$$
 or, equivalently
  $$L_{{g^*}^{-1}}\Pi^*(g^*)(\xi_1,\xi_2)= -(Ad_{{ \varsigma(g^*)}}\varsigma_R(\xi_2), p_L  Ad_{{\varsigma (g^*)}}\varsigma_R(\xi_1))_\D, \quad g^*\in G^*, \quad \xi_1,\xi_2\in \G.\eqno(2.3)$$
 Here $\varsigma:G^*\to D$ is the Lie group homomorphism
 integrating the inclusion map $\G^*\hookrightarrow\D$ and $p_R,p_L:\D\to \D$ are 
 projectors with the  kernel  $\varsigma(\G^*)$ and the respective images $\varsigma_R(\G_R)$ and $\varsigma_L(\G_L)$.
 The group $D$ is called the double 
 of the affine Poisson group $(G^*,\Pi^*)$.

 \medskip

   \noi       Note  that the projectors $p_L$,$p_R$ and their respective 
    adjoints $p_L^*$, $p_R^*$ 
 with respect to the bilinear form $(.,.)_\D$     can be all viewed as   elements of  $\D\ot\D^*$.  Since the dual $\D^*$ can
    be identified with $\D$ via    $(.,.)_\D$, we can view them also as elements
    of $\D\ot \D$.  In the latter case we denote them as $P_L,P_R,P^*_L$ and $P^*_R$,
    respectively. Obviously, $(P_L-P_L^*)$ and $(P_R-P_R^*)$ are in $\D\w \D$.

    \medskip

  \noindent{\bf Lemma 4}
  The following bivector on the group manifold $D$ is Poisson:
  $$\Pi_D=\jp L_*(P_L-P_L^*)+\jp R_*(P_R-P_R^*).\eqno(2.4)$$
  Moreover,   the bivector $\Pi_D$ is invertible on an open subset   $S$   of elements of $D$ which
    can be simultaneously decomposed as products $\varsigma(u)\varsigma_L(v_L)$ and $\varsigma_R(v_R)\varsigma(\ti u)$
    for some $u,\ti u\in G^*$, $v_L\in G_L$ and $v_R\in G_R$.
    
  \rightline{\#}

   \noi {\bf Remark}:  The symplectic manifold $(S,\Pi_D)$ is nothing but the so-called 
   global symplectic grupoid
   integrating  the so-called Lie algebroid that corresponds to the affine Poisson structure $(P,\Pi^*)$ (see \cite{Lu1,LuWe} for more details). We do not describe  here the grupoid
   structure of $S$ since we shall not need it in our study of the $q\to\infty$ limit of the quasitriangular
   WZW model.   However, we shall
   continue to use the term  symplectic grupoid in order not to give to the well-known  structure a new
   name.

  \section{Poisson-Lie symmetry}

  In this paper, we shall use somewhat abbreviated terminology,  by  calling 
  the Poisson-Lie symmetry of a Poisson manifold $P$ what is usually referred  to  in the literature as the
  infinitesimal Poisson-Lie symmetry with moment map (cf. \cite{Lu1,KS,DZ}). Moreover, we use the results
  of $\cite{Lu1,Lu2}$  to rewrite the definition of this concept in the following form:
  
  \medskip

  \noindent {\bf Definition 2}
  A Poisson manifold $(P,\Pi)$ is Poisson-Lie symmetric with respect to an affine Poisson
  group $(G^*,\Pi^*)$ if it exists a Poisson map $\mu:(P,\Pi)\to (G^*,\Pi^*)$.     
  
   \rightline{\#}

   \noi    If $\Pi^*$ in $e^*$
  vanishes (does not vanish), the Poisson-Lie symmetry is called equivariant (anomalous). 
  The Poisson map $\mu$ is referred to as the moment map and, as it is established in the following
  lemma, it permits  {\it to express
  infinitesimal symmetry transformations in terms of Poisson brackets on the manifold $P$}.

\noindent{\bf Lemma 5} Let $\lm$ ($\rho$) be the left (right) invariant Maurer-Cartan form on the
group manifold $G^*$.  If  $\mu:(P,\Pi)\to (G^*,\Pi^*)$ is a Poisson map, then the section
 $\Pi(.,\mu^*\lm)\in \G^*\ot TP$ realizes 
 a  left  $\G_L$ action on  $P$  and the section $\Pi(\mu^*\rho,.)\in \G^*\ot TP$ a right $\G_R$ action on $P$.
 
 \rightline{\#}

 \noi We observe that the Poisson-Lie symmetric manifolds always admit
 the {\it simultaneous} actions of {\it two symmetry  Lie algebras}. However,  in the equivariant case, the Lie
 algebras $\G_L$ and $\G_R$ are necessarily  isomorphic and, moreover, if the affine Poisson group $G^*$ is
 Abelian, even the $\G_L$ and $\G_R$ actions on $P$ coincide.

\medskip

 \noindent {\bf Definition 3}
 We say that  the double $D$   of 
an affine Poisson group $(G^*,\Pi^*)$ is proper  if the images of the group homomorphisms
$\varsigma,\varsigma_L,\varsigma_R$ are all simply connected and if the unit $e_D$ of $D$ is the
unique element of the intersection $\varsigma_L(G_L)\cap \varsigma(G^*)$ and also of the intersection $\varsigma_R(G_R)\cap\varsigma(G^*)$.   

 \rightline{\#}

 \noi  The properness of a double $D$ means that every element $K\in S\subset D$
 can be unambiguously decomposed in two ways:  as  $K=\varsigma_R(v_R)\varsigma(u)$, $v_R\in G_R$,
 $u\in G^*$ and as $K=\varsigma(\ti u)\varsigma_L(v_L)$, $v_L\in G_L$, $\ti u\in G^*$. These decompositions
 obviously define  four  maps 
$\Lm_L:S\to G^*$, $\Lm_R:S\to G^*$,  $\Xi_R:S\to G_L$, $\Xi_L:S\to G_R$ as follows:
$$\Lm_L(K)\equiv \ti u,\quad \Lm_R(K)\equiv  u^{-1},\quad \Xi_L(K)= v_R,\quad \Xi_R(K)=v_L^{-1}.\eqno(3.1)$$
 
\noindent{\bf Theorem 1} The maps $\Lm_L:(S,\Pi_D)\to (G^*,\Pi^*_{op})$ and $\Lm_R:(S,\Pi_D)\to (G^*,\Pi^*)$ are both
Poisson. 
\rightline{\#}

\medskip

\noindent{\bf Proof:}
 We have to show that
 $$ \Lm_{L*}\Pi_D(K)=\Pi^*_{op}(\Lm_L(K)).\nn$$
 First we note that $p_L+p_L^*=p_R+p_R^*\in \D\ot\D^*$ is nothing but the identity map from $\D$ to $\D$.
 It then follows that $P_L+P_L^*=P_R+P_R^*\in \D\ot\D$ is  $Ad$-invariant since it is the dual of the bilinear
 form $(.,.)_\D$. Thus from the equality  $L_*(P_L+P_L^*)=R_*(P_R+P_R^*)$ we deduce
 $$ \Pi_D=L_* P_L-R_* P_R^*.\nn$$
 From the very definition of the map $\Lm_L$ it follows that
 $$\Lm_{L*}R_{\varsigma_L(v_L)*}w= \Lm_{L*}w,    \quad \Lm_{L*} L_{\varsigma(u)*}w=L_{u*}\Lm_{L*}w\eqno(3.2)$$
 for a whatever vector $w\in T_KS$ and whatever elements $v_L\in G_L$ and $u\in G^*$. (Here e.g. $L_{K*}$ means
 the left transport by the element $K\in D$.) By using the relations (3.2), we easily arrive at
 $$\Lm_{L*}L_{K*}P_L=L_{\Lm_L(K)*}\Lm_{L*} Ad_{\varsigma_L(\Xi_R^{-1}(K))}P_L, \ 
 \Lm_{L*}R_{K*}P_R^*= L_{\Lm_L(K)*} \Lm_{L*}Ad_{\varsigma(\Lm_L^{-1}(K))}P_R^*.\eqno(3.3)$$
  With the help of Eqs. (3.3), we infer for any $\xi_1,\xi_2\in \G$
 $$ <\Lm_{L*}\Pi_D(K), L^*_{\Lm_L(K)^{-1}}(\xi_1\ot \xi_2)>=
 <\Lm_{L*}\biggl(Ad_{\varsigma_L(\Xi_R^{-1}(K))}P_L-Ad_{\varsigma(\Lm_L^{-1}(K))}P_R^*\biggr), \xi_1\ot \xi_2>=\nn$$
 $$
 = 0 -(Ad_{{ \varsigma(\Lm_L(K))}}\varsigma_L(\xi_2), p_R  Ad_{{\varsigma (\Lm_L(K))}}\varsigma_L(\xi_1))_\D=
 < \Pi^*_{op}(\Lm_L(K)), L^*_{\Lm_L(K)^{-1}}(\xi_1\ot \xi_2)> .\nn$$
 In a similar manner, we show that
  $$ \Lm_{R*}\Pi_D(K)=\Pi^*(\Lm_R(K)).\nn$$

\rightline{\#}

 \noi {\bf Remark}:      We note
 that our study of the structure of the symplectic grupoids of affine Poisson groups is
 similar in spirit to the study  
 of the symplectic groupoids of Poisson-Lie groups in \cite{AM}.

 \medskip

 \noi Because the moment map $\Lm_R$  is Poisson with respect to the affine Poisson structure
 $\Pi^*$, it simultaneously realizes 
 the right $\G_R$ Poisson-Lie symmetry 
 and the left $\G_L$ Poisson-Lie symmetries. However,  because the moment map $\Lm_L$ is Poisson with respect to the {\it  opposed} affine Poisson structure $\Pi_{op}^*$, it simultaneously realizes the 
  {\it left} $\G_R$ Poisson-Lie symmetry  and the {\it right} $\G_L$ Poisson-Lie symmetry.  In the applications of our general theory, presented
 in the next chapter, we shall need the explicit formulae for the right Poisson-Lie symmetries of $S$ induced by the moment maps $\Lm_L$ and $\Lm_R$. 
 
 \medskip

 \noi{\bf Theorem 2}
 The section $\Pi(\Lm_L^*\rho,.)$ generates
the infinitesimal version of the natural right $G_R$ action on $D$: $(g_R,K)\to \varsigma_R(g_R^{-1})K$
for $g_R\in G_R$ and $K\in D$. Similarly, the section  $\Pi(\Lm_R^*\rho,.)$ 
generates  
the infinitesimal version of the natural  right $G_L$ action on $D$: $(g_R,K)\to K \varsigma_L(g_L)$
for $g_L\in G_L$ and $K\in D$.

\medskip

 \noindent{\bf Proof:} Consider a point $K\in S$ and elements $\ti \eta\in\D^*$, $\xi\in \G$.  We set  $\rho_\xi\equiv<\rho,\xi>=R^*\xi$ and write 
$$ <\Pi_D, \Lm_L^*\rho_\xi \ot R_{K^{-1}}^*\ti\eta>=  
<L_{K*}P_L-R_{K*}P^*_R, \Lm_L^*\rho_\xi \ot R_{K^{-1}}^*\ti\eta>=\nn$$
$$=<P_L, (R_{\Lm_L(K)^{-1}}\Lm_L L_K)^*\xi\ot ( L_KR_{K^{-1}})^*\ti\eta>
-<P_R^*, (R_{\Lm_L(K)^{-1}}\Lm_L R_K )^*\xi \ot  \ti\eta>. \nn$$
Denote by $\eta$ the element of $\D$ which corresponds to $\ti\eta\in \D^*$ upon the identification
by the bilinear form $(.,.)_\D$.  Then
$$ <\Pi_D, \Lm_L^*\rho_\xi \ot R_{K^{-1}}^*\ti\eta>  =<(R_{\Lm_L(K)^{-1}}\Lm_L L_K)^*\xi, p_LAd_K\eta>
-<(R_{\Lm_L(K)^{-1}}\Lm_L L_K)^*\xi,p_R^*\eta>.\nn$$
Taking into account Eqs. (3.3), we infer
$$ <\Pi_D, \Lm_L^*\rho_\xi \ot R_{K^{-1}}^*\ti\eta>  =\nn$$$$
=
(\varsigma_R(\xi), Ad_{\varsigma(\Lm_L(K))}p_L^*\biggl(Ad_{\varsigma_L(\Xi_R(K)^{-1})}p_L Ad_K\eta 
 -Ad_{\varsigma(\Lm_L(K)^{-1})}p_R^*  \eta\biggr))_\D=\nn$$
$$ = 
  - (\varsigma_R(\xi),p_R^*\eta)_\D= -<\ti\eta,\varsigma_R(\xi)>
=-<R_{K*}\varsigma_R(\xi),R^*_{K^{-1}}\ti\eta>.\nn$$
We thus arrive to the announced conclusion
$$\Pi_D( \Lm_L^*\rho_\xi,.)=-R_{*}\varsigma_R(\xi).\eqno(3.4)$$
In a similar manner, we show that
$$\Pi_D(\Lm_R^*\rho_\xi,.)= L_{*}\varsigma_L(\xi).\eqno(3.5)$$
\rightline{\#}

\noi In Section 4, we shall need an explicit formula for the symplectic form $\om_D$  corresponding
to the Poisson bivector $\Pi_D$ on a {\it proper} double $D$.

\medskip

\noi{\bf Theorem 3} Let $(D,\Pi_D)$ 
be a proper double   and $S\subset D$ be the
corresponding symplectic grupoid. Denote  by $\rho$, $\rho_{L}$ and
$\rho_{R}$ the right-invariant  Maurer-Cartan forms on the respective  groups
$G^*$,$G_L$ and $G_R$.   The symplectic form $\om_S$ on $S$ is then given
by the following formula 
 $$\om_S=\jp(\Lm_L^*\rho \st \Xi^*_L \rho_{R})_\D +\jp (\Lm_R^*\rho \st \Xi^*_R \rho_{L})_\D.\eqno(3.6)$$
 
 \medskip

\noindent{\bf Proof:} Choose a basis $t_i$ of $\G^*$ and the basis  $T^i_L $ of $\G_L$  and $T_R^i$ of $\G_R$  
such that
$$(t_i,T^j_L)=\delta_i^j,\quad (t_i,T^j_R)=\delta_i^j.\eqno(3.7)$$
 The form $\om_S$ can be then rewritten as  
 $$ \om_S=\jp(\Lm_L^*\rho,T_R^i)_\D\w( \Xi^*_L\rho_{R},t_i)_\D +
 \jp (\Lm_R^*\rho,T^i_L)_\D\w( \Xi^*_R \rho_{L},t_i)_\D.\nn$$
 We are going to show that the $2$-form $\om_S$ is  the  inverse of the  Poisson bivector $\Pi_D$  restricted to $S$.

  \medskip

  \noindent  Consider a point $K\in S$ and four linear subspaces
 of the tangent space $T_KS$ defined as $S_L=L_{K*}\G_L$, $S_R=R_{K*}\G_R$,
 $\tilde S_L=L_{K*}\G^*$ and  $\tilde S_R=R_{K*}\G^*$.
At every $K\in S$ (but not necessarily at every $K\in D$!) the tangent space
$T_KS$ can be decomposed as $T_KS=S_L+\ti S_R$ and $T_KP=\ti S_L +S_R$, respectively.
We introduce a projector $\Pi_{L\ti R}$ on $\ti S_R$ with a
kernel $S_L$, a projector $\Pi_{\ti L R}$ on $S_R$ with a
kernel $\ti S_L$, a projector $\Pi_{R\ti R}$ on $\ti S_R$ with a
kernel $S_R$ and a projector $\Pi_{\ti LL}$ on $S_L$ with a
kernel $\ti S_L$. Note that the first subscript stands for the kernel and the second for the image.
 Then we have 
$$ <(\Lm_L^*\rho,T^i_R)_\D, t>=(R_{K*}T^i_R,\Pi_{L\ti R}t)_\D,\eqno(3.8)$$
 $$ <(\Xi_L^*\rho_{R},t_i)_\D, t>=(R_{K*}t_i,\Pi_{\ti L R}t)_\D,\eqno(3.9)$$
 $$ <(\Lm_R^*\rho,T^i_L)_\D, t>=-(L_{K*}T^i_L,\Pi_{R\ti L}t)_\D,\eqno(3.10)$$
 $$ <(\Xi_R^*\rho_{L},t_i)_\D, t>=-(L_{K*}t_i,\Pi_{\ti R L}t)_\D,\eqno(3.11)$$
 where  $t$ is a vector at a point $K\in S$.

 \medskip

\noi  Let us show how to demonstrate (3.8-11) on an example (3.8). For $K\in S$,
 the vectors $L_{K*}T^i_L$, $R_{K*}t_i$ form the basis of the tangent space
 $T_KS$. Thus it is sufficient to prove (3.8) for $t$ being one of the elements of the basis of $T_KS$.
For   $t=L_{K*}T^j_L$, it is obvious that the r.h.s. of (3.8) vanishes. On the other hand, knowing that 
$\Lm_L(Ke^{sT^j_L})=\Lm_L(K)$, we can  evaluate  the l.h.s.:
$$ <(\Lm_L^*\rho,T^i_R)_\D, L_{K*}T^j_L>=<(\rho,T^i_R)_\D,\Lm_{L*}(L_{K*}T^j_L)> =0.\nn$$
For $t=R_{K*}t_j$, the r.h.s. of (3.8) gives 
$$ (R_{K*}T^i_R,\Pi_{L\ti R} R_{K*}t_j)_\D=(R_{K*}T^i_R, R_{K*}t_j)_\D=\delta^i_j.\nn$$
On the other hand, knowing that $\Lm_L(e^{st_j}K)=e^{st_j}\Lm_L(K)$, we can evaluate the l.h.s.:
$$<(\Lm_L^*\rho,T^i_R)_\D, R_{K*}t_j>=<(\rho,T^i_R)_\D,\Lm_{L*}R_{K*}t_j> =\nn$$
 $$ =<(\rho,T^i_R)_\D, R_{\Lm_L(K)*}t_j>= (R_{\Lm_L^{-1}(K)*}R_{\Lm_L(K)*}t_j,T^i_R)_\D=(t_j,T^i_R)_\D=\delta^i_j.\nn$$
By using the relations (3.8-11), we can evaluate the form $\om_S$ on   any two vectors $t,u\in T_KP$
in terms of the projectors:
$$2\om_{S}(t,u)=\nn$$
$$ =(R_{K*}T^i_R,\Pi_{L\ti R}t)_\D(R_{K*}t_i,\Pi_{\ti LR}u)_\D
+(L_{K*}T^i_L,\Pi_{R\ti L}t)_\D(L_{K*}t_i,\Pi_{\ti RL}u)_\D-(t\leftrightarrow u)=\nn$$
 $$=(\Pi_{L\ti R}t,\Pi_{\ti LR}u)_\D+(\Pi_{R\ti L}t,\Pi_{\ti RL}u)_\D -(t\leftrightarrow u).\nn$$
 Here $(.,.)_{\D}$ is the bi-invariant
metric at the point $K$.
By realizing that it holds
$$ (t,\Pi_{\ti LR}u)_\D=(\Pi_{R\ti L}t,\Pi_{\ti LR}u)_\D=(\Pi_{R\ti L}t,u)_\D,\nn$$
$$\Pi_{\ti LR}+\Pi_{R\ti L}=Id,\nn$$
we  finally arrive at
$$\om_S(t,u)=(t,(\Pi_{\ti LR}-\Pi_{L\ti R})u)_\D.\nn$$
Now we can easily  show that $\om_S$ 
is the symplectic form corresponding to the Poisson structure  $\Pi_D$ restricted to $S$. First
of all,  we remark that  $\Pi_D$ can be written also as
$$ \Pi_D=L_{*}(T^i_L\ot t_i)- R_{*}(t_i\ot T^i_R).\nn$$
 Then we conclude
$$\Pi_D(.,\om_S(.,u))=\nn$$$$=L_{K*}T^i_L(L_{K*}t_i,(\Pi_{\ti LR}-\Pi_{L\ti R})u)_{\D}-
R_{K*}t_i(R_{K*}T^i_R,(\Pi_{\ti LR}-\Pi_{L\ti R})u)_{\D}=\nn$$
$$ =(\Pi_{\ti LL}-\Pi_{R\ti R})(\Pi_{\ti LR}-\Pi_{L\ti R})u=(\Pi_{R\ti R}\Pi_{L\ti R}- \Pi_{\ti LL}\Pi_{L\ti R} +
\Pi_{\ti L L}\Pi_{\ti L R})u=\nn$$
$$=(\Pi_{L\ti R} - \Pi_{\ti LL}\Pi_{L\ti R} +\Pi_{\ti L L})u=
 (\Pi_{L\ti R}+\Pi_{\ti RL})u+(\Pi_{\ti LL}-\Pi_{\ti R L} -
\Pi_{\ti L L}\Pi_{ L \ti R})u=\nn$$
$$= (\Pi_{L\ti R}+\Pi_{\ti RL})u=u. \eqno(3.12)$$
From the equation (3.12), we learn that the form $\om_S$ is invertible and its inverse is nothing
but the  Poisson  bivector $\Pi_D$ restricted to $S$. From this
it also follows, by the way,  that $\om_S$ is closed hence symplectic.

 \rightline{\#}

\section{$q\to\infty$ WZW model on a   compact group}

\subsection{$q\to\infty$ limit of a twisted Heisenberg double}

\noi We remind that  the quasitriangular WZW model (or the $q$-WZW model for short) 
is the $q$-deformation of the standard WZW model. It exhibits the Poisson-Lie symmetries with respect to two different (chiral) actions
of the polynomial loop group $L_{pol}K$ on the phase space of the model and those symmetries become
Hamiltonian in the limit $q\to 1$.  We believe that this paper would become too long if we attempted
to review here the full structure of the $q$-WZW model, and, in particular, the details of the limit $q\to 1$.Thus, for the sake of economy, we
shall point out only few facts concerning the finite $q$ that are indispensable for the good understanding of the $q\to\infty$
limit.  The reader, who will feel a need to learn more about the situation for the finite $q$, can consult
\cite{K04,K05}.

\medskip

\noi The symplectic structure of the  quasitriangular WZW model  for finite $q$ is that of the twisted Heisenberg double of the complexified  polynomial loop group $L_{pol}K^\bc$ \cite{K05}.  The concept of  the twisted Heisenberg double $(D,\k)$ is due to Semenov-Tian-Shansky \cite{ST2} and it can be defined
for every automorphism $\k$ of the double $D$ of a Poisson-Lie group $G^*$  such that $\k$
 preserves the canonical invariant bilinear form $(.,.)_\D$ on 
$Lie(D)$.   Thus $(D,\k)$ is a Poisson manifold with the Poisson bivector $\Pi_D^\k$ defined as
follows
$$\Pi_D^\k= \jp L_* (P -P^*) + \jp R_* \k_*(P-P^*).\eqno(4.1)$$
In this formula, the elements $P,P^*\in \D\w\D$ are defined as in  Lemma 4, but we abandon
the subscripts $L,R$ in $P_L,P_R$ since for the Poisson-Lie group $P_L=P_R\equiv P$.

\medskip

 \noi {\bf Remark.}  The  
 notions of the symplectic grupoid and of the twisted Heisenberg double  are not quite equivalent,
 inspite of certain similarity  between their respective  Poisson structures   (2.4)
 and  (4.1). However, in some circumstances they become equivalent; e.g. if
 the automorphism $\k$  preserves the subgroup $\varsigma(G^*)$. In this special case,
 the twisted Heisenberg double  $(D,\k)$ of a Poisson-Lie group $G^*$  can be interpreted as the symplectic grupoid
 of certain $\k$-dependent  affine Poisson structure on $G^*$ for which  the original Poisson-Lie structure
 on $G^*$ is the left-associated Poisson-Lie structure. This particular situation takes place for 
 the finite $q$  quasitriangular WZW  model,  however, as we shall see, in the limit $q\to\infty$ this is no more
 the case and the usage of the symplectic grupoids instead  of  the   twisted Heisenberg doubles becomes essential.

 \medskip

  \medskip

  \noi    Let  us  view  a   connected simply connected simple compact  Lie group $K$   
  as a subgroup of the group of  unitary matrices of order $n$.
 Following \cite{PS}, the group of polynomial loops
  $L_{pol}K$ consists of  matrix valued functions $\gamma(\si)\in K$, defined on the circle parametrized
  by an angular variable $\si$,   for which there  exist non-negative integers $p_+,p_-$ and $n\times n$-matrices $\gamma_k$
such that 
  $$\gamma(\si) = \sum_{k=-p_-}^{k=p_+}\gamma_k e^{ik\si} .\eqno(4.2)$$
Denote by $(.,.)_\K$ the negative-definite
$Ad$-invariant Killing form on the Lie algebra $\K\equiv Lie(K)$  
 and define a  non-degenerate $Ad$-invariant bilinear form
 $(.\vert .)$ on $ Lie (L_{pol}K)$ by the following formula
$$(\al\ve \bt)={1\over 2\pi}\int_{-\pi}^\pi d\si (\al(\si),\bt(\si))_{\K}.  \eqno(4.3)
 $$
 Denote by $D$  the complexification $\lpkc$ of $L_{pol}K$ and  view it as the {\it real}  group. We   note that the
 elements of $D$ are also of the form (4.2), however, for each $\si$, the result of the summation is in $K^\bc$ and not just in $K$.
 Let  $G^*$ be a subgroup of $D$, consisting  of the  elements of $\lpkc$ of the form (4.2), for which $p_-=0$
 and $\gamma_0\in AN$. Here $AN$ is the subgroup of $K^\bc$ defined by the Iwasawa decomposition $K^\bc =KAN$.
We remark  that  the elements of  $G^*$ can be viewed as the boundary values of the holomorphic maps
$\ti\gamma:\{z\in\bc:\vert z\vert \le 1\}\to K^\bc$. The following factorization lemma was proved in \cite{PS} and
it has  a crucial importance for the present paper:

\noindent{\bf Lemma 6} Any element $l\in D$ can be factorized uniquely as $l=uv$,  $u\in G^*$ and $v\in \lpk$ and
the product map $G^*\times \lpk\to D$ is a diffeomorphism.
\rightline{\#}
\noi {\bf Remark.} In what follows, we shall suppress the symbols $\varsigma$,$\varsigma_L$ and
$\varsigma_R$, standing for the group homomorphisms introduced in Sec 2.  We do it in order to
avoid cumbersome formulae and we hope that the reader will easily reconstitute them from the
context.

\medskip

 \noi Let us introduce a
nondegenerate bilinear form   $(.,.)_{\D}$ on $\D\equiv Lie(D)$  defined as 
$$(x,y)_\D={1\over \ep} Im (x\vert y), \quad x,y\in \D.\eqno(4.4)$$
 Here $(.\ve .)$ is just the bilinear form (4.3) naturally extended to $Lie(\lpkc)$ , $Im$ stands for the imaginary part    and $\ep\equiv {\rm ln}q$ is the deformation parameter.  It is not difficult to establish that
 the restrictions of the bilinear form $(.,.)_\D$ on  $Lie(\lpk)$ and on $Lie(G^*)$ both vanish. 
 For $Lie(\lpk)$, it follows from the
fact that  the bilinear form $(.\ve .)$ is real
 when restricted to $Lie(\lpk)$ and for $Lie(G^*)$, it follows from the
fact that the integral of the product of two functions containing only non-negative Fourier modes
vanishes unless both functions contain a zero mode. However,   the zero
mode part of $Lie(G^*)$ is $Lie(AN)$  and the restriction of   $(.,.)_\D$ on $Lie(AN)$ vanishes
(cf. Sec 4.4.1 of \cite{K04}).

 \medskip

 \noi The  data $D=\lpkc$, $G^*$, $\lpk$ and $(.,)_\D$, that we have just introduced, define a Poisson-Lie structure
 on $G^*$. This follows from Lemma 3 and Eqs. (2.2), (2.3), for the special case when $\G_L=\G_R=Lie(\lpk)$.  Let $\k$
 denote the following automorphism $\k$ of $D$, preserving the bilinear form $(.,.)_\D$:
$$\k(l)(\si)=l(\si+ik\ep),\quad l\in \lpkc.\eqno(4.5)$$
 The double $(D,\k)$ is then nothing but the twisted Heisenberg double of the Poisson-Lie
 group $G^*$ and the Poisson bivector (4.1) determines the symplectic structure of the quasitriangular WZW model
 for finite $q$. Note also that the
 integer parameter $k$, appearing in (4.5), becomes the level
of  the standard WZW model in the limit $q\to 1$.

\medskip

\noi In what follows, it will be useful to work with an  appropriate choice of the basis of
 $\D=Lie(\lpkc)$. We normalize the extension of the Killing-Cartan form $(.,.)_\K$ on 
$K^\bc$   in such a way that the square of the length of the longest root is 
equal to two. We pick an orthonormal Hermitian
basis $H^\mu$ in the Cartan subalgebra 
$\H^\bc$ of $\K^\bc$ with respect to 
the Killing Cartan
form   $(.,.)_\K$.  
Consider the root space decomposition of $\K^\bc$:
$$ \K^\bc=\H^\bc\bigoplus(\oplus_{\al\in\Phi}\bc E^\al),\nn$$
where $\al$ runs over the space $\Phi$ of all roots 
$\al\in\H^{*\bc}$. The step generators $E^\al$
fulfil
$$ [H^\mu,E^\al]=\al(H^\mu)E^\al,\quad(E^\al)^\dagger=
E^{-\al};\nn$$
$$  [E^\al,E^{-\al}]=\al^{\vee},\quad 
[\al^\vee,E^{\pm\al}]=\pm 2E^{\pm\al},\quad 
(E^\al,E^{-\al})_{\K}=
{2\over \vert \al\vert^2}.\nn$$
The element $\al^\vee\in \H^\bc$ is called the coroot of
 the root $\al$.
Thus the (ordinary Cartan-Weyl) basis of the complex Lie algebra 
$\K^\bc$ is $(H^\mu,E^\al)$, $\al\in\Phi$.
The affine Cartan-Weyl basis of $Lie(\lpkc)$ is now formed  by the 
elements
   of the form  
$$ E^\al e^{in\si}\equiv E^\al_n, \quad n\in\bz, 
\qquad H^\mu e^{in\si}\equiv H^\mu_n, \quad n\in\bz.\nn$$
 We call the elements $E^\al_n,H^\mu_n$
  the affine step generators with the exception of $H^\mu_0$ which 
will be called the affine Cartan generators. The commutation relations in the affine
Cartan-Weyl basis easily follow from those of the ordinary Cartan-Weyl basis 
of $\K^\bc$.  The automorphism $\k$ of $D$ defined by  (4.5) descends
   to the following automorphism of $Lie(D)$:
   $$\k_*E^\al_n=q^{-nk}E^\al_n, \quad n\in\bz, 
\qquad \k_*H^\mu_n=q^{-nk}H^\mu_n, \quad n\in\bz.\eqno(4.6)$$
In what follows, we shall
often denote   a generic affine step generator as $E^{\hat\al}$, 
where $\hal\in\hat\Phi$ stands
for the corresponding labels $(\al,n)$ or $(\mu,n\neq 0)$.  If $\hal$
is such that $\al,\mu$ are   arbitrary and $n>0$, or $\al >0$
 and $n=0$, 
we say that $\hal >0$.

\medskip

\noi A basis of the Lie subalgebra 
$Lie(\lpk)\subset Lie(\lpkc)$
can be then chosen as $(T_L^\mu,B_L^{\hal},C_L^{\hal})$,
 $\hat\al>0$ where 
$$T_L^\mu=iH^\mu,\quad B_L^{\hal}={i\over \od}(E^{\hal}+E^{-\hal}),\quad 
C_L^{\hal}={1\over \od}(E^{\hal}-E^{-\hal}).\eqno(4.7)$$
Here by $-\hal$ we mean $(-\al,-n)$ for $\hal=(\al,n)$ and  
 $(\mu,-n)$ for $\hal=(\mu,n)$. The  meaning of the subscript  $L$ will become clear soon.

\medskip

\noi 
In a similar manner, a basis  of $Lie(G^*)$ will be denoted as
$(t_\mu,b_{\hal},c_{\hal})$, $\hal >0$ and it reads:
$$ t_\mu=H^\mu,\quad b_{\hal}={\ve \hal\ve^2\over \od}E^{\hal},\quad 
c_{\hal}=-i{\ve \hal\ve^2\over \od}E^{\hal}.\nn$$
Note that for the roots of the type $\hal=(\mu,n)$, we set $\vert\hal\vert^2=2$.

\medskip

\noi  With the help of the respective  basis of $Lie(\lpk)$ and of $Lie(G^*)$,  the  Poisson structure (4.1)
of the quasitriangular WZW model for a finite $q$
can be rewritten as
$$  \Pi_D^\kappa= 
 L_*\biggl(T^\mu_L\ot {t_\mu}+B^{\hal}_L\ot {b_{\hal}}+
C^{\hal}_L\ot c_{\al}\biggr)
  -  R_*\k_*\biggl(t_\mu \ot T^\mu_L
 +b_{\hal}\ot B^{\hal}_L+
  c_{\hal}\ot C^{\hal}_L\biggr).\nn$$
  In this expression, the limit $q\to\infty$ can be directly performed with the help 
  of Eqs. (4.6) and (4.7). The result is as follows
  $$ \Pi_D^\infty= 
 L_*\biggl(T^\mu_L\ot {t_\mu}+B^{\hal}_L\ot {b_{\hal}}+
C^{\hal}_L\ot c_{\al}\biggr)
  -  R_*\biggl(t_\mu \ot T^\mu_R
 +b_{\hal}\ot B^{\hal}_R+
  c_{\hal}\ot C^{\hal}_R\biggr),\eqno(4.8)$$
  where for $n>0$ we define
  $$B_R^{\hal}={i\over \od}E^{-\hal},\quad C_R^{\hal}=-{1\over \od}E^{-\hal}, \ \hal>0\nn$$
  and for $n=0$
  $$  T_R^\mu=iH^\mu,\quad B_R^{(\al,0)}={i\over \od}(E^{\al}_0+E^{-\al}_0),\quad 
C_R^{(\al,0)}={1\over \od}(E^{\al}_0-E^{-\al}_0), \ \al>0.\nn$$
With  the help of the theory of the affine Poisson groups, it is easy to understand the structure of the
Poisson bivector $\Pi_D^\infty$. In fact, it turns out  that it defines a symplectic structure of the symplectic
grupoid $S_\infty$ corresponding to certain affine Poisson structure on the group $G^*$.
To see that, first we introduce a subgroup $G_R$ of $D$,  consisting  of the  elements of $\lpkc$ of the form (4.2), for which $p_+=0$
 and $\gamma_0\in K$.  
We remark  that  the elements of  $G_R$ can be viewed as the boundary values of the holomorphic maps
$\ti\gamma:\{z\in\bc\cup\infty:\vert z\vert \ge 1\}\to K^\bc$.  It is not difficult to see that the elements    $(T_R^\mu,B_R^{\hal},C_R^{\hal})$,
 $\hat\al>0$ form the basis of the Lie algebra 
$Lie(G_R)$. The  data $D=\lpkc$, $G^*$, $G_L\equiv \lpk$ and $G_R$  define the affine Poisson structure
 on $G^*$, given by Eqs. (2.2),(2.3).  Finally, by working out the  corresponding bivector (2.4) in the chosen basis
 on $\G_L\equiv Lie(\lpk)$, $\G_R\equiv Lie(G_R)$ and $\G^*\equiv Lie(G^*)$, we obtain the  formula (4.8).

 \subsection{$q\to \infty$   current algebras} 

  So far we have worked out the $q\to\infty$ limit of the Poisson structure of the $q$-WZW
  model and we have established that it coincides with the Poisson structure of the symplectic grupoid  
  of certain affine Poisson structure on the group $G^*$.
  In order to establish the Poisson-Lie symmetries
  of this symplectic grupoid, we wish  to show that the double $D$ is proper in the sense of Definition 3.
  This is indeed the case because, first of all, the subgroups $G_L$, $G_R$ and $G^*$ are all simply connected (cf. \cite{PS}), secondy, because $G_L$ intersects $G^*$ only at  the unit element $e_D\in D$ (cf. Lemma 6) and, finally, because
  any common element of $G_R$ and $G^*$  defines a  global
   holomorphic map from  the Riemann sphere into $K^\bc$ and, therefore, it must be $\si$-independent
   element  of  $D$ belonging at the same time to  $K$  and to $AN$. Obviously, only such element is again $e_D$. 
   We  thus see that the hypothesis of Theorem 1  are satisfied and the  $q\to\infty$ limit
   of the $q$-WZW model
enjoys a rich structure of the Poisson-Lie symmetries defined with the help
of the moment maps $\Lm_L$, $\Lm_R$ defined on the phase space $S_\infty\equiv G_RG^*\cap G_LG^*\subset D$.     The most interesting feature of this symmetry 
structure   is the fact      that the right and the left symmetry Lie algebras $\G_R$ and $\G_L$ are not
isomorphic. This is a genuinely new phenomenon arising in the $q\to\infty$ WZW model which has no analogue for
any finite $q$ and, as we shall see, it is in the origin of an interesting duality of the chiral version of the $q\to\infty$ model.

\medskip

\noi Before studying the issue  of the $\G_L\leftrightarrow \G_R$ duality, we have to work out  several
general formulae in our loop group context, in particular we shall need the affine Poisson
structures $\Pi^*$ and $\Pi^*_{op}$  on the group $G^*$. The bivector $\Pi^*$, with respect to which the map $\Lm_R$
is Poisson, is given by the general formula (2.2), while its opposed bivector $\Pi^*_{op}$, with respect to which the map $\Lm_L$
is Poisson, is given by the same
formula with the subscripts $L$ and $R$ interchanged.  As it is costumary in the studies of WZW models, we shall
characterize the Poisson bivectors $\Pi^*$ and $\Pi^*_{op}$  by Poisson brackets of a set of particular coordinate
functions on the group $G^*$ called the "Kac-Moody functions". Actually, these Kac-Moody functions, pull-backed by the
moment maps $\Lm_L$ and $\Lm_R$ to the phase space $S_\infty=G_RG^*\cap G^*G_L=G_RG^*$, are the observables
of the WZW model that generate the symmetries via the Poisson brackets on $S_\infty$, and they are commonly referred
to as the "Kac-Moody" currents.  Actually, for $q=1$, the Poisson brackets of the Kac-Moody currents give the standard
current algebra  commutators; for a finite $q$, they give the $q$-current algebra relations 
described in \cite{K04}
and for $q\to\infty$ they give $\infty$-current algebra relations that we are going to work  out.

\medskip

\noi  
The Kac-Moody  functions on the group $G^*$, the Poisson brackets of which determine
 the bivectors $\Pi^*$ and $\Pi^*_{op}$, are obviously the same as for the finite $q$. To describe them,
  we pick an irreducible
unitary representation $\Upsilon$ of the compact
   group $K$ and consider it as the representation of the complexified group $K^\bc$. We pick
   also a point on the loop characterized by a particular value of the angle coordinate $\si$.
   The data $(\Upsilon,\si)$ define  $r\times r$ functions on $LK^\bc$, where $r$ is the dimension
   of the representation $\Upsilon$. In words, these functions are defined as follows: take an element
   $l\in \lpkc$ , consider the element $l(\si)\in K^\bc$  and, finally,  matrix
 elements $ij$
   of the element $l(\si)$ in the representation $\Upsilon$.    
   \medskip

   \noi   The functions $\Upsilon_\si^{ij}$ are holomorphic. Since we regard $G^*$ as 
the {\it real}
   group, the Poisson brackets   of just holomorphic functions
   cannot fully describe the Poisson structures $\Pi^*$ and $\Pi^*_{op}$. In fact,
   we must also consider the antiholomorhic functions $(\Upsilon_\si^\dagger)^{-1}$ and
 calculate
   their brackets with the holomorphic ones. The calculation is 
considerably simplified  if
   one  uses the notation of
the matrix valued Poisson brackets \cite{FT,K04}. Thus, if $V$ is a
 vector
   space and $E$ and $F$ two $End(V)$-valued  
   functions on $G^*$, we introduce a matrix Poisson bracket  $\{E\ptp F\}$ as  the 
   $(End(V)\ot End(V))$-valued  
   function on $G^*$ defined as 
$$\{E\ptp F\}^{ik,jl}=\{E^{ij},F^{kl}\}.\nn$$
  We wish to
    calculate  Poisson brackets $\{.,.\}^*$ or $\{.,.\}^*_{op}$ 
of the matrix valued functions $\Upsilon_\si$, $(\Upsilon_\si^\dagger)^{-1}$
restricted on $G^*$.    The calculation is not difficult, it is based on  the following
obvious relations
$$<L_*t,d\Upsilon_\si^{ij}>= (\Upsilon_\si)^{ik}\Upsilon(t)^{kj};\eqno(4.9)$$
$$<R_*t,d\Upsilon_\si^{ij}>=\Upsilon(t)^{ik} \Upsilon_\si^{kj},\eqno(4.10)$$
for every $t\in\D$. The result reads:
    $$  \{\Upsilon_\si\ptp \Upsilon_{\si'}\}^*= r(\si-\si')\biggl(\Upsilon_\si \ot \Upsilon_{\si'}\biggr) -    \biggl(\Upsilon_\si \ot \Upsilon_{\si'}\biggr)r(\si-\si'). \eqno(4.11)$$
    $$  \{{\Upsilon^\dagger_\si}^{-1}\ptp {\Upsilon^\dagger_{\si'}}^{-1}\}^*=r(\si-\si')\biggl({\Upsilon^\dagger_\si}^{-1} \ot {\Upsilon^\dagger_{\si'}}^{-1}\biggr) -    \biggl({\Upsilon^\dagger_\si}^{-1}\ot {\Upsilon^\dagger_{\si'}}^{-1}\biggr)r(\si-\si'). \nn$$
 $$ \{{\Upsilon_\si^\dagger}^{-1}\ptp \Upsilon_{\si'}\}^* 
 = r(\si-\si')\biggl({\Upsilon_\si^\dagger}^{-1}\ot \Upsilon_{\si}'\biggr) -    \biggl({\Upsilon_\si^\dagger}^{-1}\ot \Upsilon_{\si}')\biggr)(r+iC).\nn$$
$$ \{\Upsilon_\si\ptp \Upsilon_{\si'}\}^*_{op}=  \{\Upsilon_\si\ptp \Upsilon_{\si'}\}^*; \quad
\{{\Upsilon^\dagger_\si}^{-1}\ptp { \Upsilon^\dagger_{\si'}}^{-1}\}^*_{op}= 
 \{{\Upsilon^\dagger_\si}^{-1}\ptp {\Upsilon^\dagger_{\si'}}^{-1}\}^*;\eqno(4.12)$$
$$  \{{\Upsilon_\si^\dagger}^{-1}\ptp \Upsilon_{\si'}\}^*_{op}
 = (r+iC)\biggl({\Upsilon_\si^\dagger}^{-1}\ot \Upsilon_{\si}'\biggr) -    
 \biggl({\Upsilon_\si^\dagger}^{-1}\ot \Upsilon_{\si}')\biggr)r(\si-\si').\nn$$
    Here the canonical $r$-matrix  and the Casimir element $C$ are given by the standard expressions
   $$ r=\sum_{\al>0}{i\vert\al\vert^2\over 2}(E^{-\al}\ot E^\al-E^\al\ot
E^{-\al}),\nn$$
 $$ C=\sum_\mu H^\mu\ot H^\mu +\sum_{\al>0}{\vert\al\vert^2\over 2}
(E^{-\al}
\ot E^\al+E^{\al}\ot E^{-\al}),\nn$$
 moreover, we set
 $$ r(\si-\si') = i\sum_\mu(H^\mu\ot H^\mu) (1+2\sum_{ n>0} 
   e^{in(\si-\si')})+$$ $$\quad{}
    +i\sum_{\al>0}\ve\al\ve^2(E^{-\al}\ot E^\al)+
   i\sum_{\al}\ve\al\ve^2(E^{-\al}\ot E^\al)\sum_{ n>0} 
   e^{in(\si-\si')} 
  =r+C{\rm cotg}{\si-\si'\over 2}.\eqno(4.13)$$
  For completeness, we evaluate explicitely also  
   the left and the right Poisson-Lie brackets $\{.,.\}^*_L$ and
 $\{.,.\}^*_R$
  associated to the affine Poisson structure $\Pi^*$,  by using
    the defining formulae (2.2) and (2.3)  with all subscripts $L,R$ set to $L$ only, for $\{.,.\}^*_L$,
    and to 
  $R$ only,  for  $\{.,.\}^*_R$:
    $$  \{\Upsilon_\si\ptp \Upsilon_{\si'}\}^*_L=\{\Upsilon_\si\ptp \Upsilon_{\si'}\}^*_R=  \{\Upsilon_\si\ptp \Upsilon_{\si'}\}^*; \nn$$
    $$ 
\{{\Upsilon^\dagger_\si}^{-1}\ptp { \Upsilon^\dagger_{\si'}}^{-1}\}^*_L
=\{{\Upsilon^\dagger_\si}^{-1}\ptp { \Upsilon^\dagger_{\si'}}^{-1}\}^*_R=
  \{{\Upsilon^\dagger_\si}^{-1}\ptp {\Upsilon^\dagger_{\si'}}^{-1}\}^*;\nn$$
$$  \{{\Upsilon_\si^\dagger}^{-1}\ptp \Upsilon_{\si'}\}^*_L
 = r(\si-\si')\biggl({\Upsilon_\si^\dagger}^{-1}\ot \Upsilon_{\si}'\biggr) -    \biggl({\Upsilon_\si^\dagger}^{-1}\ot \Upsilon_{\si}')\biggr)r(\si-\si').\nn$$
 $$  \{{\Upsilon_\si^\dagger}^{-1}\ptp \Upsilon_{\si'}\}^*_R
 = (r+iC)\biggl({\Upsilon_\si^\dagger}^{-1}\ot \Upsilon_{\si}'\biggr) -    \biggl({\Upsilon_\si^\dagger}^{-1}\ot \Upsilon_{\si}')\biggr)(r+iC).\nn$$
 Let us define the following Hermitian matrix valued observables on the phase space 
  $S_\infty=G_RG^*\cap G^*G_L$  of the $q\to\infty$ WZW model:
  $$ L(\si) =\Lm_L^*(\Upsilon_\si\Upsilon_{\si}^\dagger )
   ,\quad R(\si)=\Lm_R^*(\Upsilon_{\si}^\dagger \Upsilon_\si).\nn$$
  By using the Poisson properties of the maps   $\Lm_L:(S_\infty,\Pi_D^\infty)\to (G^*,\Pi^*_{op})$ and $\Lm_R:(S_\infty,\Pi_D^\infty)\to (G^*,\Pi^*)$, we find out the basic commutation relations of the
  left and the right $\infty$-current algebras:
$$\{L(\si)\ptp L(\si')\}_D^\infty= $$
$$\quad{} =\biggl(L(\si)\ot L(\si')\biggr)\biggl(r+C{\rm cotg}{\si-\si'\over 2}\biggr)+
  \biggl(r+C{\rm cotg}{\si-\si'\over 2}\biggr)\biggl(L(\si)\ot L(\si')\biggr)$$
$$\quad{}  -\biggl(L(\si)\ot 1 \biggr)\biggl(r+iC\biggr)
  \biggl(1\ot L(\si')\biggr)- \biggl(1\ot L(\si')\biggr)\biggl(r-iC\biggr)\biggl(L(\si)\ot 1 \biggr),\eqno(4.14)$$
    $$\{R(\si)\ptp R(\si')\}_D^\infty=$$
$$\quad{} =-\biggl(R(\si)\ot R(\si')\biggr)\biggl(r+C{\rm cotg}{\si-\si'\over 2}\biggr)-
  \biggl(r+C{\rm cotg}{\si-\si'\over 2}\biggr)\biggl(R(\si)\ot R(\si')\biggr)+$$
$$\quad{} +\biggl(R(\si)\ot 1 \biggr)\biggl(r-iC\biggr)
  \biggl(1\ot R(\si')\biggr)+ \biggl(1\ot R(\si')\biggr)\biggl(r+iC\biggr)\biggl(R(\si)\ot 1 \biggr).\eqno(4.15)$$
  The Poisson brackets (4.14) and (4.15)  are the $q\to\infty$ analogues of the ordinary
  $q\to 1$ left and right Kac-Moody relations (2.66) and (2.67) of Ref. \cite{K04}.
  Actually, for a generic finite $q$,   we have  concentrated ourselves 
in \cite{K04} mainly on the left chiral version of 
  the $q$-WZW model
 therefore we have detailed in (1.37) of \cite{K04}
 only the left $q$-Kac-Moody  brackets.
  It is in fact the  bracket (1.37) of \cite{K04} 
which   is the finite $q$  analogue of our left $\infty$-bracket (4.33).   

 \subsection{Chiral decomposition}
 Both the standard WZW model \cite{W,Gaw,Mad} and its $q$-deformation \cite{K04}  admit the so called
 chiral decomposition. This  means, roughly speaking,
 that the phase space $S_q$ of the model for each finite $q$ can 
 be represented as a "square" of a simpler symplectic manifold $M_q$ which itself
 enjoys
 only one half of the full symmetry  of $S_q$.  More precisely, by the "square" 
of a symplectic manifold $M_q$ we mean the symplectic manifold
 $S_q=M_q\times M_q//T$ where the notation $//T$ means the symplectic reduction
 by an appropriate  action of the Cartan torus $T\subset K$ on $M_q\times M_q$. The aim of the section is to show
 that  the chiral decomposition takes place also in the $q\to\infty$ case. However, there is a novelty:
  a remarkable duality in the description
 of  the chiral symplectic
 manifold $M_\infty$ related to the fact that the groups $G_R$ and $G_L$ are not isomorphic.
 We shall describe this duality in Sec 4.6 and, for the moment, we restrict ourselves
 to the explicite description of the chiral decomposition of the phase space $S_\infty=G_RG^*\cap G^*G_L=
 G_RG^*$.
 Let us start by formulating and proving an auxiliary theorem on $\infty$-Cartan decomposition.
 We remind that, in order to avoid cumbersome notations, we shall not write explicitely the injection
 homomorphisms $\varsigma$,
 hoping that the reader will restore them easily from the context.

\vskip1pc

\noi{\bf Theorem 4}  {\it   For very element $s\in S_\infty$,
  there exist two elements $k_l,k_r\in G_R$
 and an element $a\in A_+$  such that
 $$s=k_la\Xi_R(k_ra).\eqno(4.16)$$
 The ambiguity of this decomposition is given by the simultaneous right multiplication   $(k_l,k_r)\to
 (k_lt,k_rt)$
  by any element $t$ of the Cartan torus $T\subset K$. 

\medskip

\noi Reciprocally, for every elements $k_l,k_r\in G_R$
 and every  element $a\in A_+$ the product  $k_la\Xi_R(k_ra)$ is in $S_\infty$.}

 \vskip1pc

\noindent{\bf Proof:}
First of all we remind the notation.  Thus  $A$ is the (real) subgroup of $K^\bc$
 given by the Iwasawa decomposition $K^\bc=KAN$.  Its Lie algebra $\A$  consists of the Hermitian
 elements of the Cartan subalgebra $\H^\bc$ and it is spanned by the elements $H^\mu$ (cf. Sec 4.1).
 By $A_+$ we mean  $\exp{\A_+}$ where $\A_+$ is the positive Weyl chamber in $\A$. 
 For instance, for the groups $K^\bc=SL(n,\bc)$, $A_+$ consists of diagonal matrices with
 real positive entries ordered from the biggest one to the smallest one.  We remind also that
 the Lie algebra $\T$ of the Cartan torus $T$ is spanned by the elements $T^\mu=iH^\mu$.
 Finally, we recall that the decomposition $D=G^*G_L$ is global  hence the domain of definition
 of the map $\Xi_R$ (introduced in Eq. (3.1)), is the whole  double $D$.

 \medskip

 \noi The (hermitian conjugated version of the) theorem 8.1.1. of \cite{PS} says that
 every element $u\in\lpkc$ can be decomposed as $u=u_-u_0$ where the
 Fourier expansion (4.2) of $u_-$  contains only the non-positive modes and 
  $u_0$ is in the subgroup of $\lpk$ 
  consisting of the loops passing  through the unit element of $K$ at $\si=0$.
  Obviously, $u$ can be decomposed also as $u=u_Nu'u_0$. Here $u_0$ is as before, $u'$
  is in $K^\bc$ ( in this context  $K^\bc$ is viewed as the subgroup of $LK^\bc$ formed by the constant
  loops) and the Fourier expansion $u_N$ contains  as before only the non-positive
  modes with the zero mode being equal 
  to the unit element of $K^\bc$. By the classical theorem about the Cartan decomposition
  of a simple complex connected and simply connected  group $K^\bc$ (cf. \cite{Zel}, p. 117), we may write $u'=u_lau_r^{-1}$, where $u_l,u_r$ are the elements of $K$ and $a$ is in $A_+$.
 The ambiguity of this classical Cartan decomposition is given by the simultaneous right multiplication   $(u_l,u_r)\to
 (u_lt,u_rt)$
  by any element $t$ of the Cartan torus $T$. Thus we may decompose $u$ also as
  $$ u=u_Nu_l au_r^{-1}u_0.\nn$$
  We note that $u_Nu_l$ is an element of the subgroup $G_R$ of $\lpkc$ defined at the end of
   Sec 4.1 and $u_r^{-1}u_0$
  is the element of the subgroup $G_L=\lpk$. For the moment, we conclude that any $u$ in $\lpkc$
  can be decomposed as
  $$u=k_lag_r^{-1},\nn$$
  where $k_l\in G_R$, $a\in A_+$ and  $g_r\in G_L$. The ambiguity of this decomposition 
  is given by the simultaneous right multiplication   $(k_l,g_r)\to
 (k_lt,g_rt)$
  by any element $t$ of the Cartan torus $T$.

  \medskip 

  \noi Suppose now, that the element $u$ is in $S_\infty=G_RG^*$. This means that the 
  element $ag_r^{-1}$ is in the domain of definition of the maps $\Xi_L$ and $\Lm_R$
  (cf. Eq. (3.1)).  Set
  $$ k_r=\biggl(\Xi_L(ag_r^{-1})\biggr)^{-1}\equiv\Xi_L^{-1}(ag_r^{-1}).\nn$$
  We see that $k_r$ is an element of $G_R$.  Since the domain of definition
 of the map $\Xi_R$ is the whole  double $D$, we can evaluate $\Xi_R(k_ra)\in G_L$:
 $$  \Xi_R(k_ra)= \Xi_R(\Xi_L^{-1}(ag_r^{-1})a) 
   = \Xi_R(\Lm_R(ag_r^{-1})\Xi_L^{-1}(ag_r^{-1})a)=
 \Xi_R(g_ra^{-1}a)=g_r^{-1}.\nn$$
 We conclude that for every $u\in S_\infty$ there exist $k_l,k_r\in G_R$ and $a\in A_+$ such
 that
 $$ u=k_la\Xi_R(k_ra).\nn$$
 It remains to deal with the ambiguity of this decomposition.
 When $g_r$ is replaced by 
 $g_rt$, $t\in T$ then $k_r$ is replaced by 
$\Xi_L^{-1}(t^{-1}ag_r^{-1})=\Xi_L^{-1}(ag_r^{-1})t=k_rt$.
This proves the first part of the theorem.

\medskip

\noi Reciprocally, let $k_l,k_r$ be in $G_R$ and $a\in A_+$. In order to show
that $k_la\Xi_R(k_ra)$ is in $S_\infty=G_RG^*$, it is obviously sufficient to show that
$a\Xi_R(k_ra)$ is in $S_\infty$.
We use the fact that the maps $\Xi_R$ and $\Lm_L$ are defined everywhere on $D=\lpkc$
and we write
$$ a\Xi_R(k_ra)=a\Xi_R(k_ra)\Lm_L^{-1}(k_ra)\Lm_L(k_ra)=a(k_ra)^{-1}\Lm_L(k_ra)=
k_r^{-1}\Lm_L(k_ra).\nn$$
The  element $k_r^{-1}\Lm_L(k_ra)$ is evidently in $G_RG^*$.

\rightline{\#}

  \noi We have just established that the phase space $S_\infty$ can be identified
with the manifold
$(G_R\times A_+\times G_R)/T$. The core of this section
 is the following theorem expressing the chiral decomposability
 of the $q\to \infty$ WZW model:
 
 \vskip1pc
 
 \noi{\bf Theorem 5}
 Parametrize by $k_l\in G_R$, $a_l\in A_+$ 
  the direct product $M_\infty \equiv G_R\times A_+$ and
 define the following $2$-form $\Om_\infty$ on $M_\infty$:
 $$ \Om_\infty(k_l,a_l)=\jp(da_la^{-1}_l\st k_l^{-1}dk_l)_\D+\jp ( d\Xi_R(k_la_l)\Xi_R^{-1}(k_la_l)\st
 a^{-1}_lda_l+a^{-1}_l(k_l^{-1}dk_l ) a_l)_\D.\eqno(4.17)$$
  Denote by $\phi:G_R\times A_+\times G_R\to S_\infty$ the map induced by the
$\infty$-Cartan decomposition, i.e.
$$\phi(k_l,a,k_r)= k_la\Xi_R(k_ra).\eqno(4.18)$$
Then the pull-back $\phi^*\om_{S_\infty}$ of the grupoid symplectic form (3.6)
can be written as
$$ \phi^*\om_{S_\infty}= \Om_\infty(k_l,a_l=a)-\Om_\infty(k_r,a_r=a).\nn$$
 
 \medskip

\noindent{\bf Proof:}
The theorem says, in other words,  that the restriction of
the form $\Om^l_\infty-\Om_\infty^r$ on the submanifold  
of $M_\infty\times M_\infty$  defined
by $a_l=a_r$ is the same thing as the $\phi^*$-pull-back of the anomalous Semenov-Tian-Shansky 
form from $S_\infty$ into $G_R\times A_+\times G_R$. In order to prove it let $s$
denote an element of $S_\infty$. The grupoid symplectic form
 $\om_{S_\infty}$ is given by Eq.(3.6)
$$\om_{P_\infty}=\jp(d\Lm_L(s)\Lm_L^{-1}(s)\st d\Xi_L(s)\Xi_L^{-1}(s))_\D+
\jp(d\Lm_R(s)\Lm_R^{-1}(s)\st d\Xi_R(s)\Xi_R^{-1}(s))_\D.\eqno(4.19)$$
By using the $\infty$-Cartan parametrization (4.16)
 and the relations (3.1), we easily infer:
$$\Lm_L(s)=\Lm_L(k_la)=k_la\Xi_R(k_la),\eqno(4.20)$$
$$\Lm_R(s)=\Lm_R(a\Xi_R(k_ra))=\Xi_R^{-1}(k_ra)a^{-1}k_r^{-1},\eqno(4.21)$$
$$\Xi_L(s)=k_l\Xi_L(a\Xi_R(k_ra))=k_lk_r^{-1},\eqno(4.22)$$
$$\Xi_R(s) =\Xi_R^{-1}(k_ra)\Xi_R(k_la).\eqno(4.23)$$
By inserting the expressions (4.20-23) into (4.19), we find immediately
$$\phi^*\om_{S_\infty}= 
 \jp(daa^{-1}\st k_l^{-1}dk_l)_\D+\jp ( d\Xi_R(k_la)\Xi_R^{-1}(k_la)\st
 a^{-1}da+a^{-1}(k_l^{-1}dk_l ) a)_\D$$ $$ \quad{}
  -\jp(daa^{-1}\st k_r^{-1}dk_r)_\D-\jp ( d\Xi_R(k_ra)\Xi_R^{-1}(k_ra)\st
 a^{-1}da+a^{-1}(k_r^{-1}dk_r ) a)_\D=$$ $$ \quad{}
  =\Om_\infty(k_l,a_l=a)-\Om_\infty(k_r,a_r=a).\nonumber$$

\rightline{\#}

\noi {\bf Corollary} {\it  The  chiral form $\Om_\infty$ is symplectic.  The manifold 
$S_\infty$
can be obtained by the symplectic reduction $S_\infty=M_\infty\times M_\infty//T$,
 where
$T$ is the Cartan torus acting as $(k_l,k_r)\to (k_lt,k_rt)$, $t\in T$.}

\medskip

\noindent{\bf Proof:} 
The closedness of $\Om_\infty$ can be seen from the fact that  $\Om_\infty(k_l,a_l)$ 
 is the pull-back of the closed form $\om_{S_\infty}$ under the
 map $\chi:G_R\times A_+\to S_\infty$ given by $\chi(k_l,a_l)= k_la_l$.  It 
 is slightly more involved to
show the non-degeneracy of $\Om_\infty$ .  First of all we compute the contraction
$\iota_v\Om_\infty$ where $v=L_{k_l*}T^\mu$. The result is
$$\Om_\infty(.,L_{k_l*}T^\mu)= (T^\mu, da_la_l^{-1})_\D= d\psi_l^\mu,\eqno(4.24)$$
where  we have parametrized $a_l$ as $a_l=e^{\psi_l^\mu H^\mu}$. 
We observe that, whatever
  is the point  $(k_l,a_l)$ in $M_\infty$,
 the vector $L_{k_l*}T^\mu$ does not constitute
a degeneracy direction of the form $\Om_\infty$. If the form $\Om_\infty$ had at some point
a degeneracy vector, then the restriction of the form $\Om_\infty(k_l,a_l)-\Om_\infty(k_r,a_r)$
to the submanifold $a_l=a_r$ would have  degeneracy vectors {\it other}  than $L_{k_l*}T^\mu + L_{k_l*}T^\mu$. However, this is impossible since $\Om_\infty(k_l,a_l)-\Om_\infty(k_r,a_r)$
is the pull-back of the non-degenerate form $\om_{P_\infty}$. 

\medskip

\noi Having established that $(M_\infty,\Om_\infty)$ is a symplectic manifold
(actually it is going to play the role
of the phase space of the chiral $q\to\infty$ WZW model), it is easy to prove that
$S_\infty =M_\infty\times M_\infty//T$.  Indeed, due to relative minus sign of the left and the
right chiral symplectic forms on each copy of $M_\infty$, we see from (4.24) that the Hamiltonian
function $\psi^\mu_l-\psi^\mu_r$  generates the action of the vector  $L_{k_l*}T^\mu 
+ L_{k_r*}T^\mu$
on $M_\infty\times M_\infty$. The setting $a_l=a_r$ is nothing but saying  that $\psi^\mu_l-\psi^\mu_r=0$ 
and we conclude that the anomalous Semenov-Tian-Shansky form
$\om_{S_\infty}$ comes indeed from the symplectic reduction 
$M_\infty\times M_\infty//T$.

\rightline{\#}

\medskip

\noi {\bf Remark.}   The proof  of the corollary is not rigorous  by strictly mathematical standards
and  shifts the paper to the level of rigour common in mathematical physics. Indeed,
the propositions in  Sections 2 and 3 were proved only for  finite dimensional
Lie groups, in particular the result about the non-degeneracy  of the grupoid
symplectic form (3.6).  In our study of the infinite-dimensional loop groups, we still preserve
the full mathematical rigour for certain propositions, e.g. Theorems 4 and 5, but  in some
cases we adopt the usual cavalier approach of mathematical physicists in treating the
infinitesimal symplectic geometry of field theories. Thus, in the proof of the corollary, we assumed
that the theorems of Sections 2 and 3 remains valid also in the infinite dimensional context.

\subsection{Symmetry of the chiral model}
There is a natural right action of the group $G_R$ on the chiral phase space $M_\infty$
given by $h\tr (k_l,a_l)=(h^{-1}k_l, a_l)$, $h\in G_R, (k_l,a_l)\in M_\infty$. It  can be interpreted
as the restriction to $M_\infty\subset S_\infty$ of 
   the  Poisson-Lie symmetric right action   $G_R\times S_\infty\to S_\infty$ given
by the group multiplication on the anomalous double: $h^{-1}\tr s= hs$, $h\in G_R,
 s\in S_\infty$.  
Our next goal is to prove that also  the restricted (or chiral)
$G_R$-action is in fact a Poisson-Lie symmetry.

\vskip1pc

\noi{\bf Theorem 6}
 Denote 
$\Pi^\infty$   the bivector inverse to the chiral symplectic form $\Om_\infty$ and consider
the map $\chi:G_R\times A_+\to S_\infty$ given by $\chi(k_l,a_l)= k_la_l$. Then
it holds that the composition map $\Lm_L\circ \chi: (M_\infty,\Pi^\infty)\to (G^*,\Pi^*_{op})$ is Poisson.

\medskip

\noindent{\bf Proof:}
Consider   a pair of functions $x,y$ defined on $G^*$,
 their  pull-backs $\Lm_L^*x,\Lm_L^*y$ defined on $S_\infty$ and 
$\chi^*\Lm_L^*x$, $\chi^*\Lm_L^*y$ defined on $M_\infty$,
and also the functions 
$\chi^*\Lm_L^*x\ot 1, \chi^*\Lm_L^*y\ot 1$ defined on $M_\infty\times M_\infty$.
Consider  the submanifold $O$ of $M_\infty\times M_\infty$ defined by
 setting $a_l=a_r$ and also the map $\phi: O\to S_\infty$ defined by Eq. (4.18). 
   Then the $\infty$-Cartan parametrization   (4.16)
and the definition (3.1)  of the map $\Lm_L$ imply that 
$$(\chi^*\Lm_L^*x\ot 1)\vert_O=\phi^*\Lm_L^*x.\eqno(4.25)$$
Stated in words, the restriction of $\chi^*\Lm_L^*x\ot 1$ to $O$ is the same thing 
as the $\phi^*$-pull-back
of the function $\Lm_L^*x$ defined on $S_\infty$. Moreover, we have
$$ \Lm_L(k_lta_l)=\Lm_L(k_la_lt)=\Lm_L(k_la_l), \quad k_l\in G_R,a_l\in A_+, t\in T.\nn$$
This relation implies that the function $\chi^*\Lm_L^*x\ot 1$ is  $T$-invariant with
respect to the Cartan torus action defined in the Corollary of Theorem 5.   Since we know
that $S_\infty=(M_\infty\times M_\infty)//T$, we infer
$$ \{\chi^*\Lm_L^*x\ot 1, \chi^*\Lm_L^*y\ot 1\}_{M_\infty\times M_\infty}\vert_O=
\phi^*\{\Lm_L^*x,\Lm_L^*y\}_D^\infty,\nn$$
or, equivalently,
$$(\{\chi^*\Lm_L^*x, \chi^*\Lm_L^*y\}_{M_\infty}\ot 1)\vert_O=
\phi^*\{\Lm_L^*x,\Lm_L^*y\}_D^\infty.\eqno(4.26)$$
If we write the identity (4.25) for the function $\{x,y\}^*_{op}$, Theorem 1 combined with Eq. (4.26)
gives
$$ (\{\chi^*\Lm_L^*x, \chi^*\Lm_L^*y\}_{M_\infty}\ot 1)\vert_O=
 (\chi^*\Lm_L^*\{x,y\}^*_{op}\ot 1)\vert_O.\nn$$
 Thus we conclude 
$$\{\chi^*\Lm_L^*x, \chi^*\Lm_L^*y\}_{M_\infty}=
 \chi^*\Lm_L^*\{x,y\}^*_{op}.\nn$$

\rightline{\#}

\medskip

\noi {\bf Corollary}  {\it The right action of the group $G_R$ on the $\infty$-WZW  chiral phase space $M_\infty$,
given by $h\tr (k,a)= (h^{-1}k, a)$, $h\in G_R$,  is the right  
Poisson-Lie symmetry corresponding to the moment map $\Lm_L\circ \chi$. }

\medskip

\noindent{\bf Proof:}
Consider a point $(k,a)$ in $M_\infty$.  
 The multiplication of $(k,a)$ on the left by an infinitesimal
generator $S\in \G_R$ gives the vector 
$v_{S}=(Sk,a)$ and its $\chi_*$-push-forward vector
 $\chi_*v_S= Ska=R_{(ka)*}S\in T_{(ka)}S_\infty$. Denote also by $v_S$  and $\chi_*v_S$
the respective  vector fields on $M_\infty$ and $S_\infty$ obtained by varying the point $(k,a)$.
In the sense of Lemma 5, Theorem  1 and Eq. (3.4) of Sec 4.3,
we know that  
$$-\chi_*v_S=\Pi^\infty_D(\Lm_L^*(\rho,S)_\D,.).\nn$$
This relation can be rewritten equivalently as
$$\om_{S_\infty}(.,\chi_*v_S)= \Lm_L^*(\rho,S)_\D\eqno(4.27)$$
(note that the bivector inverse to the groupoid symplectic  form $\om_{S_\infty}$
on $S_\infty$ has been denoted by $\Pi^\infty_D$).
The $\chi^*$-pull-back of Eq. (4.27) gives
$$\Om_{\infty}(.,v_S)= \chi^*\Lm_L^*(\rho,S)_\D,\eqno(4.28)$$
or, equivalently,
$$ v_S=-\Pi^\infty((\Lm_L\circ\chi)^*(\rho,S)_\D,.).\eqno(4.29)$$

\rightline{\#}

 \subsection{Exchange relations}
 Our next goal is to invert  the chiral symplectic form $\Om_\infty$ on $M_\infty =G_R\times A_+$. The strategy, which we shall use, will be that of Sec 5.1.2 of \cite{K04}. It consists in exploiting
 particular properties of the Lie derivatives of Poisson-Lie symmetric  bivectors. Let $u$ be a differential $1$-form on $M_\infty$ and
 $v=R_*S$, $S\in\G_R$ a right-invariant   vector field on $G_R$ viewed as the vector field on $M_\infty$.  We calculate the Lie derivate
 $\L_v$ of the both sides of the following identity
 $$\Om_\infty(.,\Pi^\infty(.,u))=u\nn$$
 and we obtain
 $$ (\L_v\Om_\infty)(.,\Pi^\infty(.,u))+\Om_\infty(.,(\L_v\Pi^\infty)(.,u))=0.\nn$$
 From 
 Eq. (4.28), we infer
 $$\L_v\Om_\infty= d(\iota_v\Om_\infty)=-d( \chi^*\Lm_L^*(\rho,S)_\D))= 
 - \chi^*\Lm_L^*((\rho,S')_\D\w (\rho,S'')_\D),\eqno(4.30)$$
 where the map $S\to S'\w S''\in \G_R\w\G_R$ is the dual to the Lie algebra commutator $[.,.]^*:\G^*\w\G^*\to\G^*$. The last equality follows from
 the well-known Maurer-Cartan identity
 $$d(R^*S)=R^*S'\w R^*S'', \quad S\in (\G^*)^*,\eqno(4.31)$$
 because $\G_R$ can be identified  with  $(\G^*)^*$.
Finally,  by inserting (4.30) into (4.31), we arrive at the following formula
$$\L_v\Pi^\infty=-v'\w v",\nn$$
where $v'\equiv R_*S'$ and $v''\equiv R_*S''$.

\medskip

\noi With the notations introduced after Lemma 3, introduce
 the  following Poisson-Lie bivector $\Pi_R$ on the group manifold $G_R$:
$$R_{g^{-1}*}\Pi_R(g)(t_1,t_2)\equiv-(Ad_{\varsigma_R(g^{-1})}\varsigma(t_1),p_R^*Ad_{\varsigma_R(g^{-1})}\varsigma(t_2))_\D, \quad g\in G_R, t_1,t_2\in \G^*
\eqno(4.32)$$
 and consider
it as the bivector on $M_\infty$. Then a straightforward calculation gives 
$$\L_v(\Pi^\infty+\Pi_{R})=0,\nn$$
hence  $\Pi^\infty+\Pi_{R}$ must have the following shape
$$\Pi^\infty+\Pi_{R}= \Sigma_{ij}(a)L_*(T^i_R \w T^j_R)+
\si^\mu_i(a) L_*T^i_R\w {\partial\over \partial\phi^\mu}+
s^{\mu\nu}(a) {\partial\over \partial\phi^\mu}\w  {\partial\over \partial\phi^\nu},\nn$$
where $a=\exp{(\sum \phi^\mu H^\mu)}$ and $T^i_R$ is a basis in $\G_R$.
 The crucial observation is as follows: 
since at the group unit $e_{R}\in G_R$ the Poisson-Lie bivector $\Pi_{R}$
vanishes,  in order
to fully determine the unknown functions $ \Sigma_{ij}(a)$,$\si^\mu_i(a)$
 and $s^{\mu\nu}(a)$, it is sufficient to invert the
chiral symplectic form $\Om_\infty$ just at the points $(e_{R},a)$.  

\medskip

\noi We may parametrize a small vicinity of the  group origin $e_{R}$ by  the coordinates on the
Lie algebra $\G_R$ corresponding to the basis $(T^\mu, B_R^{\hal},C_R^{\hal})$,$\hal>0$ introduced
in Sec 4.1. Thus
we parametrize any element $\zeta\in\G_R$ as
$$\zeta=\tau_\mu T^\mu+ \beta_{\hal}B_R^{\hal}+\gamma_{\hal}C_R^{\hal}\nn$$
and find the following expression for the symplectic form $\Om_\infty$ in $(e_{R},a)$
$$\Om_\infty(e_{G_R},a)=d\phi^\mu\w d\tau_\mu+
\sum_{\hal>0,n>0}{a^{2\al}\over \vert\hal\vert^2}d\beta_{\hal}\w d\gamma_{\hal}
+ \sum_{\hal>0,n=0}{a^{2\al}-1\over \vert\hal\vert^2}d\beta_{\hal}\w d\gamma_{\hal}.\eqno(4.33)$$
Here we use the notation
$$a^{2\al}\equiv e^{2\al(\phi^\mu H^\mu)}.\nn$$
It is easy to invert this almost Darboux-like
 expression (4.33) and to write  (everywhere
in $M_\infty$):
$$\Pi^\infty= -\Pi_{G_R}+L_*T^\mu \w { \partial\over \partial \phi^\mu}
- \sum_{\hal>0,n>0}a^{-2\al} \vert\hal\vert^2 L_*(B^{\hal}_R  \w  C^{\hal}_R)
- \sum_{\hal>0,n=0}{\vert\hal\vert^2 \over a^{2\alpha}-1}
L_* (B^{\hal}_R \w   C^{\hal}_R).\eqno(4.34)$$
We wish to calculate the brackets of the principal variables
$a\in A_+$ and $k\in G_R$ like $\{a\ptp k(\si)\}_\infty$, 
 $\{k(\si)\ptp k(\si')\}_\infty$, $\{k(\si)\ptp {k(\si')^\dagger}^{-1}\}_\infty$, etc.
  or more precisely, the brackets
 $\{a\ptp \Upsilon_\si\}_\infty$,  $\{\Upsilon_\si\ptp \Upsilon_{\si'}\}_\infty$, 
 $\{\Upsilon_\si\ptp {\Upsilon^\dagger_{\si'}}^{-1}\}_\infty$ etc. where 
 the matrix valued functions $\Upsilon_\si$ (introduced in Sec 4.2) are restricted to
 the subgroup $G_R\subset \lpkc$. However, we prefer to make explicit slightly 
modified  Poisson brackets
like $\{a\ptp k(\si)a\}_\infty$, 
 $\{k(\si)a\ptp k(\si')a\}_\infty$, 
$\{k(\si)\ptp {(k(\si')a)^\dagger}^{-1}\}_\infty$, since they 
contain equivalent information
 and   are less cumbersome. With the help of Eqs.(4.32),(4.34),
 (4.9) and (4.10),  we find
$$\{a\ptp k(\si)a\}_\infty=-i(a\ot k(\si)a)(H^\mu\ot H^\mu),\eqno(4.35)$$
$$ \{k(\si)a\ptp k(\si')a\}_\infty =(r+C{\rm cotg}{\si-\si'\over 2})\times $$
$$\quad{} 
\times \biggl(k(\si)a\ot k(\si')a\biggr)-\biggl(k(\si)a\ot 
k(\si')a\biggr)
(r(a)+C{\rm cotg}{\si-\si'\over 2}),\eqno(4.36)$$
$$ \{{(k(\si)a)^\dagger}^{-1} \ptp {(k(\si')a)^\dagger}^{-1}\}_\infty=
 (r+C{\rm cotg}{\si-\si'\over 2})\times $$
 $$
 \quad{} \times \biggl({(k(\si)a)^\dagger}^{-1} 
\ot {(k(\si')a)^\dagger}^{-1}\biggr)-\biggl({(k(\si)a)^\dagger}^{-1} 
\ot {(k(\si')a)^\dagger}^{-1}\biggr)(r(a)+C{\rm cotg}{\si-\si'\over 2}),\eqno(4.37)$$
$$ \{k(\si)a\ptp {(k(\si')a)^\dagger}^{-1}\}_\infty=$$
$$\quad{}  =(r-iC)
\biggl(k(\si)a\ot {(k(\si')a)^\dagger}^{-1}\biggr)-\biggl(k(\si)a\ot 
{(k(\si')a)^\dagger}^{-1}\biggr)
(r(a)+C{\rm cotg}{\si-\si'\over 2}),\eqno(4.38)$$
where
$$r(a)=\sum_\al {i\vert \al \vert^2\over 2}{a^\al +a^{-\al}\over a^\al -a^{-\al}}
E^{-\al}\ot E^\al.\eqno(4.39)$$
Note that the summation in (4.39) runs over all roots $\al$ and not only over the
positive roots. It is not difficult to recognize in the expression (4.39) the canonical
dynamical $r$-matrix associated to a simple Lie algebra (cf. Eq.(5.60) of \cite{K04}).

\medskip

\noi The Poisson brackets (4.35-38) encode full information about the symplectic
structure of the chiral $\infty$-WZW model and they may be called the exchange
relations. However, it is not obvious how to compare them with the exchange
relations of the finite $q$ WZW model (cf. Eq.(5.159) of Ref.\cite{K04}).
The point is that the our $\infty$-exchange relations involve the Poisson brackets
of certain matrix-valued
functions on the group $G_R$ while the  exchange relations for
finite $q$ involve the Poisson brackets
of the matrix-valued
functions on the group $G_L=\lpk$. It turns out, however, that the chiral
$\infty$-WZW model enjoys a remarkable duality which enables to describe its dynamics
also in terms of the functions on the group $G_L$. Thus, in particular, we shall
see in the next section that  the exchange
relations (4.35-38) can be equivalently rewritten in terms of dual exchange
relations employing the matrix-valued functions on $G_L$.  It will turn out,
rather satisfactorily, that the dual $\infty$-exchange relation can be obtained by
the direct $q\to\infty$ limit of the finite $q$ exchange relations.

 \medskip

 \subsection{Duality}
 We have learned in Sec 3,  that  the anomalous Poisson-Lie moment maps
 realize at the same time
 the  right and the left Poisson-Lie symmetries.  In  Corollary of Theorem 6, 
 we have worked out 
  the  action of the  right infinitesimal $\G_R$ symmetry  on the phase space $M_\infty$
   of the chiral $\infty$-WZW model and 
 we found
  that it  can be lifted to  the natural global  right  action of the group $G_R$ on itself. More precisely
  it is given by $h\tr (k,a)=(h^{-1}k,a)$, $h\in G_R$, $(k,a)\in G_R\times A_+=M_\infty$.
  Consider now the  infinitesimal action of the left  
  $\G_L$ Poisson-Lie symmetry generated by 
 the vector fields $ \Pi^\infty(.,(\Lm_L\circ\chi)^*(\lm,T)_\D)$ for $T\in \G_L$    and we may ask the following question:
  Can we parametrize the points of the phase space  $M_\infty$ by using the
 group $G_L$ in such a
  way that the  action of the left  
  $\G_L$ symmetry becomes just (the infinitesimal version of) the natural 
 action of the group $G_L$ on itself?
  It turns out that the answer to this question is  affirmative, i.e. we 
shall succeed 
  to represent $M_\infty$ as a submanifold
 of the direct product $G_L\times A_-$ on
  which the Lie algebra $\G_L$ (but not the group $G_L$) acts by 
the  right-invariant
  vector fields on $G_L$.  
The quantitative basis of this result is the folowing theorem:

 \vskip1pc
 
 \noi{\bf Theorem 7}
 Parametrize by $\ti k\in G_L$, $\ti a\in A_-$  
 the direct product $G_L\times A_-$,
   where $A_-=\exp{\A_-}$ and $\A_-=-\A_+$ is 
 the negative   Weyl chambre.  
 Define a   $2$-form $ \ti\Om_\infty(\ti k,\ti a)$   on  the {\it submanifold}
  $\ti M_\infty$ of $G_L\times A_-$ for
 which $\ti a^{-1}\ti k^{-1}$ is in the domain of definition of the maps $\Xi_L$
and $\Lm_R$ (i.e. $\ti a^{-1}\ti k^{-1}$ is in $S_\infty$):
 $$ \ti\Om_\infty(\ti k,\ti a)=\jp(d\ti a\ti a^{-1}\st \ti k^{-1}d\ti k)_\D+
\jp ( d\Xi_L(\ti a^{-1}\ti k^{-1})\Xi_L^{-1}(\ti a^{-1}\ti k^{-1})\st
 \ti a^{-1}d\ti a+\ti a^{-1}(\ti k^{-1}d\ti k ) \ti a)_\D.\eqno(4.40)$$
 Then the manifold $(\ti M_\infty,\ti\Om_\infty)$
is symplectic and the
infinitesimal  version of the right  $G_L$ action  
 $\ti h\tr (\ti k,\ti a)=(\ti h^{-1}\ti k, \ti a)$, $\ti h\in G_L$  is a right 
Poisson-Lie symmetry of   $(\ti M_\infty,\ti\Om_\infty)$.

 \medskip

\noindent{\bf Proof:}
 Consider  a  map $U:M_\infty\to \ti M_\infty$ 
defined everywhere in $M_\infty$ as follows 
$$(\ti k,\ti a)=U(k,a)=(\Xi_R^{-1}(ka),a^{-1}), \quad (k,a)\in M_\infty.\eqno(4.41)$$
Similarly,  consider  a  map  $V:\ti M_\infty\to M_\infty$ defined 
everywhere in  $\ti M_\infty$ as follows  
 $$(k,a)=V(\ti k,\ti a)= (\Xi_L^{-1}(\ti a^{-1}\ti k^{-1}),\ti a^{-1}),
 \quad (\ti k,\ti a)\in \ti M_\infty.\nn$$
 We easily verify that $U\circ V$ is the identity map on $M_\infty$ and $V\circ U$ is the
 identity map  on $\ti M_\infty$. It follows that both $U,V$
 are injective and surjective, hence they are diffeomorphisms inverse to each other.

 \medskip

 \noi  By an easy direct calculation, we can relate the  form
 (4.40) on $\ti M_\infty$
 to the symplectic form (4.17) on $M_\infty$ by pull-backs
 $$U^*\ti\Om_\infty=-\Om_\infty, \quad V^*\Om_\infty=-\ti\Om_\infty.\eqno(4.42)$$
This implies that $(\ti M_\infty,\ti\Om_\infty)$ is   a symplectic manifold.

\medskip

\noi We pick  $T\in \G_L$ and we push forward by  $U$   the vector field 
$\Pi(.,\chi^*\Lm_L^*<\lm,T>)\in Vect(M_\infty)$, realizing the {\it left} $\G_L$ Poisson-Lie
symmetry of  $M_\infty$. The result is a vector field $\ti w_T\in Vect(\ti M_\infty)$ given by
 $$\ti w_T \equiv U_*(\Pi^\infty(.,\chi^*\Lm_L^*<\lm,T>))= -\ti\Pi^\infty(.,V^*\chi^*\Lm_L^*<\lm,T>)=
\ti\Pi^\infty(.,V^*\chi^*\Lm_L^*J^*<\rho,T>),\nn $$
where $\ti\Pi^\infty$
stands for the Poisson bivector inverse to the symplectic form $\ti\Om_\infty$,  the map $\chi:M_\infty\to S_\infty$ was defined in Theorem 6 and the map $J:G^*\to G^*$
is just the inversion map $J(b)=b^{-1}$, $b\in G^*$.  Note that $J^*\rho=-\lm$.

\medskip

\noi  Now consider a map $\ti\chi:\ti M_\infty\to S_\infty$ given by $\ti\chi(\ti k,\ti a)=
\ti a^{-1}\ti k^{-1}$. Then it is easy to see that
$$\Lm_R\circ \ti\chi=J\circ\Lm_L\circ\chi\circ V,\eqno(4.43)$$
which permits to rewrite
 $$\ti w_T= \ti\Pi^\infty(.,\ti\chi^*\Lm_R^*<\rho,T>). \nn$$
 Obviously, we have
 $$\ti\Om_\infty(.,\ti w_T)=\ti\chi^*\Lm_R^*(\rho,T)_\D.\eqno(4.44)$$
 Consider a point $(\ti k,\ti a)\in \ti M_\infty$. The multiplication of $(\ti k,\ti a)$ on the left by an infinitesimal
generator $T\in \G_L$ gives the vector 
$\ti v_{T}=(T\ti k,\ti a)$ and its $\ti\chi_*$-push-forward vector
 $\ti\chi_*\ti v_T= -\ti a^{-1}\ti k^{-1}T=-L_{(\ti a^{-1}\ti k^{-1})*}T\in T_{(\ti a^{-1}\ti k^{-1})*}S_\infty$. Denote also by $\ti v_T$  and $\ti\chi_*\ti v_T$
the respective  vector fields on $\ti M_\infty$ and $S_\infty$ obtained by varying the point $(\ti k,\ti a)$.
Then we learn from Theorem 2 (cf. Eq. (3.5)):
$$\ti\chi_*\ti v_T= \Pi_D^\infty(.,\Lm_R^*(\rho,T)_\D),\nn$$
or, equivalently,
$$\omega_{S_\infty}(.,\ti\chi_*\ti v_T)=\Lm_R^*(\rho,T)_\D.\eqno(4.45)$$
 The relation (4.42) and Theorem 5 imply easily
$$\ti\Omega_\infty=\ti\chi^*\om_{S_\infty},\nn$$
where $\om_{S_\infty}$ is the grupoid symplectic form (3.6).  This fact  permits to infer from (4.45):
$$\ti\Omega_\infty(.,\ti v_T)=\ti\chi^*\Lm_R^*(\rho,T)_\D.\eqno(4.46)$$
By comparing Eqs. (4.44) and (4.46), we obtain
$$\ti w_T=\ti v_T.\eqno(4.47)$$
Thus we have established, that the {\it left} $\G_L$ Poisson-Lie symmetry of the chiral phase
space $M_\infty$ is most conveniently described by using the (anti)
-symplectomorphism $U:
M_\infty\to \ti M_\infty$. Indeed, as Eq. (4.47) 
shows, in  the parametrization $\ti M_\infty= \ti G_L\times A_-$
the $\G_L$ symmetry action is nothing but  the
infinitesimal  version of the natural {\it left}  $G_L$ action  
 $\ti h\tr (\ti k,\ti a)=(\ti h\ti k, \ti a)$, $\ti h\in G_L$.

\medskip

\noi Now the proof of the present theorem is almost finished. Obviously, the {\it right} action
of $G_L$ on $\ti M_\infty$, given by  $\ti h\tr (\ti k,\ti a)=(\ti h^{-1}\ti k, \ti a)$, $\ti h\in G_L$,
is infinitesimaly generated by the vector fields $-\ti v_T$, $T\in\G_L$. We infer  from (4.44) and (4.47)
that 
$$-\ti v_T= \ti\Pi^\infty(\ti\chi^*\Lm_R^*<\rho,T>,.) =  \ti\Pi^\infty((\Lm_R\circ \ti\chi)^*<\rho,T>,.)  \nn$$
 It  remains to show that the map $(\Lm_R\circ \ti\chi):(\ti M_\infty,\ti P_\infty)\to
 (G^*,\Pi^*)$ is Poisson, thus realizing  the Poisson-Lie symmetry of $\ti M_\infty$.
 The Poisson property of the map  $\Lm_R\circ \ti\chi$ follows from Eq. (4.43). Indeed, we know from (4.42)  that 
$V:(\ti M_\infty, \ti\Pi^\infty)\to (M_\infty,-\Pi^\infty)$ is the Poisson map,  Theorem 6
states that $\Lm_L\circ\chi:(M_\infty,\Pi^\infty)\to (G^*,\Pi^*_{op})$ is the Poisson map and,
by using Eqs. (2.2),(2.3), it is not difficult to work out that $J:(G^*,\Pi^*_{op})\to (G^*,-\Pi^*)$ is the Poisson map.
Thus the composition map $J\circ\Lm_L\circ\chi\circ V=\Lm_R\circ \ti\chi$   is also Poisson. 
\rightline{\#}

\medskip

\noi{\bf Interpretation in terms of duality.} It is not so surprising that there is an 
(anti)-symplectomorphism $U$ transforming the
infinitesimal {\it left}  $\G_L$ Poisson-Lie symmetry   of  $M_\infty$
into a more nicely looking {\it right}
$\G_L$ Poisson-Lie symmetry  of $\ti M_\infty$. After all, to find this
 anti-symplectomorphism,  it was sufficient to integrate the vector fields $ \Pi^\infty(.,(\Lm_L\circ\chi)^*(\lm,T)_\D)$
 to $G_L$-orbits and
 use the $G_L$ orbits for the parametrization of the chiral phase space $M_\infty$
(it is in this way that we have constructed the diffeomorphism $U$).
  What is surprising and
 remarkable is  that the $U$-transformed expression (4.40) has {\it the same
  structure} than the original one (4.17). Indeed, we   
    have 
  $$\Om_\infty(k,a) 
  =\jp(daa^{-1}\st k^{-1}dk)_\D+\jp ( (\Xi_R\circ \chi)^*\rho_{R}\st
 a^{-1}da+a^{-1}(k^{-1}dk ) a)_\D,\nn$$
 $$ \ti\Om_\infty(\ti k,\ti a) 
 =\jp(d\ti a\ti a^{-1}\st \ti k^{-1}d\ti k)_\D+
\jp ( (\Xi_L\circ \ti\chi)^*\rho_{L}          \st
 \ti a^{-1}d\ti a+\ti a^{-1}(\ti k^{-1}d\ti k ) \ti a)_\D,\nn$$
where $\rho_{R}$ and $\rho_{L}$ are, respectively, the right-invariant
Maurer-Cartan forms on $G_R$ and $G_L$.
We interpret this phenomenon as the 
$\G_L\leftrightarrow \G_R$  duality of the chiral $q\to \infty$ WZW model.
This duality relates the  Poisson-Lie symmetries of $(M_\infty,\Om_\infty)$
and $(\ti M_\infty,\ti\Om_\infty)$ in a nice way: we have shown that 
 the left $\G_L$ symmetry 
of $(M_\infty,\Om_\infty)$ gets $U$-transformed into the  right $\G_L$ symmetry
of $(\ti M_\infty,\ti\Om_\infty)$  and it is not difficult to show that
the right $\G_R$ symmetry
of $(M_\infty,\Om_\infty)$ gets $U$-transformed into the left $\G_R$ symmetry
of $(\ti M_\infty,\ti\Om_\infty)$.

\medskip

\noi {\bf Remark.}  Corollary of Theorem 6 tells us that the right $\G_R$ Poisson-Lie  symmetry of 
$M_\infty$ is given by the
infinitesimal version of the 
 natural global right  $G_R$ action on $M_\infty=G_R\times A_+$, hence we see that the infinitesimal
 right $\G_R$ symmetry   can be lifted to the {\it  global} $G_R$ symmetry.
 However, the   left $\G_L$ Poisson-Lie  symmetry of 
$M_\infty$ {\it cannot} be lifted to 
 a global action of the group $G_L=\lpk$ on $M_\infty$. This becomes evident upon the
 $U$-transformation of $M_\infty$ into $\ti M_\infty$. Indeed, the manifold $\ti M_\infty$ is
 the {\it proper} subset of the direct product $G_L\times A_-$ and the global action of the
 group $G_L$ 
 on itself does not respect this proper subset. Thus we observe, that a moment map
 can realize at the same time a global right symmetry and a  local left symmetry. In this
 sense, the duality  $\G_L\leftrightarrow \G_R$ is only local and it cannot be lifted to a global
 $G_L\leftrightarrow G_R$ duality.

\subsection{$q\to\infty$ limit of the exchange relations}

\noi Now it is time to work out the dual exchange relations, i.e. to calculate the Poisson
brackets of the type $\{\ti a,\ti k(\si)\}_\infty$, $\{\ti k(\si)\ptp \ti k(\si')\}_\infty$.
 Actually, 
those two brackets characterize completely the  dual symplectic structure $(\ti M_\infty,\ti\Om_\infty)$.
It is because 
$\ti k$ is the element of $G_L=\lpk$ which means that ${\ti k(\si)^\dagger}^{-1}=\ti k(\si)$ and it is not necessary
to calculate the brackets $\{\ti k(\si)\ptp {\ti k(\si')^\dagger}^{-1}\}_\infty$, $\{{\ti k(\si)^\dagger}^{-1}\ptp {\ti k(\si')^\dagger}^{-1}\}_\infty$ like in the $G_R$ case.  Our strategy for working out the dual
exchange relations will be different than it was for the original relations (4.35-38).  We
shall not try to invert directly the dual symplectic  form $\ti\Om_\infty$ because
 the fact that it is not defined everywhere
on $G_L\times A_-$ makes the task more involved than the inversion of the form $\Om_\infty$ in Sec 4.5.
Instead, we shall proceed indirectly, by using  the duality 
  antisymplectomorphism $U:M_\infty\to \ti M_\infty$ and the
 Poisson-Lie symmetry of $(M_\infty,\Om_\infty)$.  Indeed, we take into account the
definition (4.41) of the diffeomorphism $U$ to see   
  that, for finding the dual brackets 
 $\{\ti a,\ti k(\si)\}_\infty$, $\{\ti k(\si)\ptp \ti k(\si')\}_\infty$, it is
sufficient to calculate  the  brackets  $\{ a^{-1},  \Xi_R^{-1}(ka)(\si)\}_\infty$,
 $\{  \Xi_R^{-1}(ka)(\si)\ptp     \Xi_R^{-1}(ka)(\si')\}_\infty$  
with the help of the original Poisson bivector (4.34).

\medskip

\noi  Consider an element $(k,a)\in G_R\times A_+=M_\infty$ and write it
as   $ka= \Lm_L(ka)\Xi_R^{-1}(ka)$. Thus we have 
$$\{ a^{-1},  \Xi_R^{-1}(ka)(\si)\}_\infty=\{ a^{-1},  
\Lm_L^{-1}(ka)(\si)k(\si)a\}_\infty,\eqno(4.48)$$
$$\{  \Xi_R^{-1}(ka)(\si)\ptp     \Xi_R^{-1}(ka)(\si')\}_\infty=
\{  \Lm_L^{-1}(ka)(\si)k(\si)a\ptp     \Lm_L^{-1}(ka)(\si')k(\si')a\}_\infty.\eqno(4.49)$$
If we take into account the following obvious Poisson matrix relations
$$ \{AB\ptp CD\} = (A\ot 1)\{B\ptp C\}(1\ot D)+ $$
$$ \quad{}+
 (A\ot C)\{B\ptp D\}  +\{A\ptp C\}(B\ot D)
+(1\ot C)\{1\ptp D\}(B\ot 1),\eqno(4.50)$$
$$\{A\ptp B^{-1}\}=-(1\ot B^{-1})\{A\ptp B\}(1\ot B^{-1}),\eqno(4.51)$$
we realize that the dual exchange relations (4.48) and (4.49) can be worked out from
the matrix Poisson brackets of the type $\{ a\ptp ka\}_\infty$,  $\{ka \ptp ka \}_\infty$, $\{\Lm_L(ka)\ptp \Lm_L(ka)\}_\infty$, $\{a\ptp \Lm_L(ka)\}_\infty$ and 
  $\{ka\ptp \Lm_L(ka)\}_\infty$.  Let us argue that we already know explicit formulae 
  for all   five brackets just listed. Indeed,  the first two brackets in the list are given in Eqs. (4.35)
  and (4.36). Next,  the fact that $\Lm_L\circ \chi:(M_\infty,\Pi^\infty)\to 
(G^*,\Pi_{op}^*)$ is the Poisson map
  implies that the third bracket is given by the  $\infty$-current algebra relation (4.11) (cf. also  (4.12)
  and (4.13)):
   $$
\{\Lm_L(ka)(\si)\ptp \Lm_L(ka)(\si')\}_\infty=\biggl(r+C{\rm cotg}{\si-\si'\over 2}\biggr)\times  $$
$$ \quad{} 
 \times  \biggl(\Lm_L(ka)(\si)\ot \Lm_L(ka)(\si')\biggr) -    \biggl(\Lm_L(ka)(\si)\ot \Lm_L(ka)(\si')\biggr) \biggl(r+C{\rm cotg}{\si-\si'\over 2}\biggr).\eqno(4.52)$$
  Finally, we notice that the    brackets $\{a\ptp \Lm_L(ka)\}_\infty$,
  $\{ka\ptp \Lm_L(ka)\}_\infty$ can be worked out  as follows:

 \medskip

\noi  
Consider  the vector field  $v_{T}$ corresponding to the action of $T\in\G_R$ on $M_\infty$.
 Corollary of Theorem 6 states that 
 $$ <v_{T},df>=\Pi^\infty(df,(\Lm_L\circ\chi)^*(\rho,T)_\D),\nn$$
 for every function $f$ defined on $M_\infty$.
  Let $\Upsilon$ be the representation of the group $K$ introduced in Sec 4.2 and
  $Tr_\Upsilon$ be   the trace in this representation  normalized
in such a way that
$$Tr_\Upsilon(\Upsilon(A)\Upsilon(B))=(A,B)_\K, \quad A,B\in\K^\bc.\nn$$
The right-invariant Maurer-Cartan form $\rho$ on $G^*$ can be written with the help
of the matrix valued functions $\Upsilon$ defined in Sec 4.2:
$$\Upsilon(\rho)=d\Upsilon\Upsilon^{-1},\eqno(4.53)$$
Thus, using (4.3) and (4.4), we can write
$$
<v_{T},df>={1\over 2\pi}\int_{-\pi}^{\pi} d\si Im Tr_\Upsilon(\Upsilon(T),\{f,  (\Lm_L\circ \chi)^*\Upsilon_\si\}_\infty 
(\Lm_L\circ \chi)^*\Upsilon^{-1}_\si).\eqno(4.54)$$
 Let $T^i_R$ a basis of $\G_R$, $t_i$ the basis of $\G^*$ verifying the condition (3.7)
and $k\in G_R$.
We wish to express the   quantity   $\Upsilon_\si(T^i_R)\Upsilon_\si(k)\ot \Upsilon_{\si'}(t_i)$, by using  the Poisson brackets on $M_\infty$.  By invoking Eqs. (4.10), (4.54) and the notation of the proof
of Corollary of Theorem 6,  we have
$$ \Upsilon_\si(T^i_R)\Upsilon_\si(k)\ot \Upsilon_{\si'}(t_i)=$$
$$=<v_{T^i_R},d\Upsilon_\si(k)>\ot \Upsilon_{\si'}(t_i)= \{\Upsilon_\si(k)\ptp \Upsilon_{\si'}(\Lm_L(ka))\}_\infty \biggl(1\ot \Upsilon_{\si'}(\Lm_L^{-1}(ka))\biggr),\nn$$
By suppressing
   the symbol $\Upsilon$, as usual, we can thus write 
   $$\{k(\si)\ptp \Lm_L(ka)(\si')\}_\infty= $$$$ \quad{}=
 (T_R^i(\si)\ot t_i(\si'))(k(\si)\ot \Lm_L(ka)(\si')) =(r+C{\rm cotg}{\si-\si'\over 2})(k(\si)\ot \Lm_L(ka)(\si')).\eqno(4.55)$$
 Note that we have evaluated the expression $T_R^i(\si)\ot t_i(\si')$ in the basis
 $T^i_R=(T_R^\mu,B_R^{\hal},C_R^{\hal})$,
 $\hat\al>0$ and $t_i=(t_\mu,b_{\hal},c_{\hal})$, $\hal >0$, defined in Sec 4.1.

 \medskip

 \noi We know that the coordinates $\phi^\mu$ are invariant with respect to the
 action of the $\G_R$ Poisson-Lie symmetry on $M_\infty$, which, combined with Eqs. (4.29) and
 (4.53),
   gives
    $$\{\phi^\mu,\Lm_L(ka)\}_\infty\Lm_L^{-1}(ka)=0.\nn$$
  From this equality we finally infer
    $$\{a\ptp \Lm_L(ka)\}_\infty= 0.\eqno(4.56)$$
  We are now ready to write down the seeken dual exchange relations. We take into account 
   Eqs. (4.35), (4.36), (4.52), (4.55), (4.56) as well as Eqs. (4.42), (4.50), (4.51) to find out
   $$\{\ti a,\ti k(\si)\}_\infty= -i(\ti a \ot \ti k(\si))(H^\mu\ot H^\mu),\eqno(4.57)$$
   $$ \{\ti k(\si)\ptp \ti k(\si')\}_\infty= $$ $$ \quad{}=
(\ti k(\si)\ot \ti k(\si')) (r(\ti a^{-1})+ C{\rm cotg}{\si-\si'\over 2})+
    (r+ C{\rm cotg}
   {\si-\si'\over 2})  (\ti k(\si)\ot \ti k(\si')).\eqno(4.58)$$
   We remind that $$\ti a=a^{-1}=e^{-\phi^\mu H^\mu}, \quad  \ti k=\Lm_L^{-1}(ka)ka\nn$$
    and the minus sign distinguishing the
   symplectic forms $\Om_\infty$ and $U^*\ti\Om_\infty$ has been taken into account. 

   \medskip

 \noindent We stress that the dual  $(\ti M_\infty, \ti\Om_\infty)$-description
  of the $q\to\infty$ chiral 
 phase space is $U$-equivalent
 to the original $(M_\infty, \Om_\infty)$-description, in particular,
 the dual $G_L$-exchange relations (4.57), (4.58) are equivalent to
 the $G_R$-exchange relations
 (4.35-38).  
  From the symmetry poin of view,   the $(M_\infty, \Om_\infty)$-formalism
 is better adapted to the explicit description  of the $\G_R$-symmetry while   the $(\ti M_\infty,\ti \Om_\infty)$-formalism
 is better adapted to the explicit description  of the
 $\G_L$-symmetry.  Note, however, one   important point: 
  the submanifold $\ti M_\infty\subset G_L\times A_-$ is not
  invariant under the 
 global $G_L$-action while the manifold $M_\infty=G_R\times A_+$ 
is invariant under the global
 $G_R$-action. Said in other words,
the $\G_R$-symmetry is global while the $\G_L$-symmetry is only local.
  This is the reason why we have constructed our exposition starting  
 from the $(\M_\infty, \Om_\infty)$-description where the   global  $G_R$-symmetry
  looks naturally.  However, the $(\ti M_\infty, \ti\Om_\infty)$-description has also its assets because, as we are going
  to see, it  is better adapted for the comparison of the finite $q$ chiral WZW model
 with the $q\to\infty$ chiral WZW model. 

 \medskip

  \noi  Now we wish to compare the dual exchange relations (4.57) and (4.58) with the exchange
  relations of the finite $q$ chiral WZW model of Ref. \cite{K04}.  To do that we should take into account
  that our parameter $q$ is the parameter $1/q$ of Ref.\cite{K04}, as it follows from the comparison
  of Eq. (4.5) of the present paper and of Eq. (4.113) of \cite{K04}. In order not to cause
  confusion, we shall denote by $q'$ the parameter $q$ of \cite{K04} and we set $\e'={\rm  ln}q'$.
  The exchange relations 
  for finite $q'$ were explicited in   Eqs. (5.159) and (5.161) of Ref. \cite{K04}  only for  $q'>1$ (or $\e'>0$) and we have to make them
  explicit also for $0<q'<1$ (since when  our $q$ goes to $\infty$ the "old" $q'$ of \cite{K04} goes to $0^+$).
  By reusing the method of Sec 5.2.4 of \cite{K04}, we obtain for   $\e' <0$:
  $$\{\ti a,\ti k(\si)\}_{\e'}= -i\e'(\ti a \ot k(\si))(H^\mu\ot H^\mu),\eqno(4.59)$$
$$ \{\ti k(\si)\ptp \ti  k(\si')\}_{\e'} = $$
$$\quad{}=
 (\ti k(\si)\otimes \ti k(\si'))
\hat r_{\e'}(\ti a,\si-\si')+
 \e'(r+ C{\rm cotg}
   {\si-\si'\over 2}) (\ti k(\si)\otimes \ti k(\si')),\eqno(4.60)$$
where $\ti k(\si)\in G_L$ and $\hat r_{\e'} (\ti a,\si)$ is the Felder 
 elliptic dynamical $r$-matrix \cite{F}
given by
 $$ \hat r_{\e'}(\ti a,\si)= $$ $$ \quad{}=
{\e'\over \pi}\rho({\si\over 2\pi},-{i k\e'\over \pi})
H^\mu\otimes H^\mu
+{\e'\over \pi}\sum_{\al\in\Phi}{\rn\over 2}\si_{-{\e' k  a^\mu\la
\al,H^\mu\ra\over \pi i}}
({\si\over 2\pi},-{i k\e'\over \pi})E^\al\otimes E^{-\al}.\nn$$
We remind that here $\e'<0$, $\ti a=e^{k\e' a^\mu H^\mu}$,
$k$ is the level of the $q$-WZW model and the coordinates $a^\mu$ parametrize the 
so called positive Weyl
alcove $\A_+^1$  (cf. Sec 5.2 of \cite{K04}).  Using the 
classical formulae  from Ref. \cite{WHW} which define  the elliptic  functions $\si_{-y}(z,\tau)$, $ \rho(z,\tau)$
$$\si_{-y}(z,\tau)=\pi({\rm cotg}\pi z+{\rm cotg}\pi y)
+ 4\pi \Sigma_{m,n>0}e^{2\pi i\tau mn}\sin{2\pi(mz+ny)};\nn$$
$$\rho(z,\tau)=\pi {\rm cotg}\pi z +
4\pi\Sigma_{n>0}{e^{2\pi in\tau}\sin{2\pi nz}\over 1-e^{2\pi in\tau}},\nn$$
  and keeping fixed the expression $\phi^\mu=- k\e' a^\mu$,
we find that 
 $$\mathop {\lim }\limits_{\e' \to -\infty }  {1\over \e'}\hat r_{\e'} (\ti a,\si)= \biggl(r(e^{\phi^\mu H^\mu})+ C{\rm cotg}{\si-\si'\over 2}\biggr),\nn$$
 This gives, in turn, for $\ti a=e^{-\phi^\mu H^\mu}$
 $$ \mathop {\lim }\limits_{\e' \to -\infty }{1\over \e'}\{\ti a,\ti k(\si)\}_{\e'}= -i(\ti a\ot k(\si))(H^\mu\ot H^\mu),\nn$$ 
$$  \mathop {\lim }\limits_{\e' \to -\infty }{1\over \e'}\{\ti k(\si),\ti k(\si')\}_{\e'}=  $$
$$ \quad{}=
(\ti k(\si)\ot \ti k(\si')) (r(\ti a^{-1})+ C{\rm cotg}{\si-\si'\over 2})+ (r+ C{\rm cotg}
   {\si-\si'\over 2})  (\ti k(\si)\ot \ti k(\si')).\nn$$
    We have just established, as   expected, that 
 $ \mathop {\lim }\limits_{\e' \to -\infty }{1\over \e'}\{., .\}_{\e'}   =  \{. \ptp .\}_\infty $, or, in other words, 
 the finite $q$ chiral WZW exchange relations (4.59), (4.60)  give in the
 $q\to\infty$ limit the (dual) $G_L$-exchange relations (4.57),(4.58) of the  chiral $\infty$-WZW model.

\medskip

\noi The reader may feel intrigued why we did not start our exposition of the $\infty$-WZW model
by first establishing the $q\to\infty$ limit of the chiral exchange relations, but, instead, we have
first exposed the theory of the affine Poisson groups etc. Actually, the more
"theoretical" approach, that we have chosen, have evident benefits. The most important
is that it allowed us to identify the singularities of the dual exchange relations (4.57) and (4.58).
 Indeed,
the following phenomenon takes place:   the chiral WZW phase space for finite $q$ is
 just the manifold $G_L\times A^1$, where $A^1$ is the (compact)  subset of $A$ obtained by 
 exponentiating  the positive  Weyl alcove as  follows: 
 $ e^{k\e' a^\mu H^\mu}\in A ^1$. In the $q\to\infty$ limit
 (or the $\e'\to -\infty$ limit), 
  the set $A^1$  gets expanded to the exponentiated negative Weyl chamber $A_-=\exp{\A_-}$  and 
 the chiral  Poisson structure,  defined  on the {\it whole}  manifold $G_L\times A_-$ 
 by the brackets (4.57) and (4.58),  becomes 
invertible  only on  the {\it submanifold} $\ti M_\infty\subset G_L\times A_-$. 
Our approach, using the theory of affine Poisson groups,  gave as the natural characterization
of the regular submanifold $\ti M_\infty$ in terms of the domain of definition
of the maps $\Xi_L$ and $\Lm_R$.

\medskip

 \noi The general  theory of the affine Poisson groups  has helped us
 to clarify one more subtle point. Indeed, as we have already mentioned, there is a price to pay for the restriction
  to the regular submanifold $\ti M_\infty\subset G_L\times A_-$, namely, the loop group
  $G_L=\lpk$ no longer acts on $\ti M_\infty$. At the first sight, this may look bad, because
  the $q\to\infty$ limit  seems to deprive  the WZW model from its interesting symmetry structure,
  however, as we have again learned from Theorems 2 and 7, 
  the  $G_L$-symmetry does survive the limit  
   in its  local $\G_L$-form.  On the top of that,
  there are  further added benifits of our approach:  we have discovered  the new $G_R$
  symmetry, which emerged in the $q\to \infty$ limit and also the remarkable 
  $\G_L\leftrightarrow \G_R$
  duality of the chiral $\infty$-WZW model.

 \subsection{$q\to\infty$ Hamiltonian}
 A dynamical system is a triple $(M,\Om,E)$ where $M$ is a phase space, $\Om$ a symplectic
 form on it  and $E$ is a one-parameter group of symplectomorphisms of $M$ defining
 the time evolution.   We have often
 spoken about a particular dynamical system called the chiral $\infty$-WZW model but, so far, 
 we have  specified only its phase space $M_\infty= G_R \times A_+$  and
 its symplectic form $\Om_\infty$ given by  Eq. (4.17).  In this section, we shall fill the
 gap and define also the one-parameter group $E_\infty$ describing  the time
 evolution in  the chiral $\infty$-WZW model.    In order to do that we  first define
 a suitable parametrization of the phase space $M_q$,   for
  a finite $q$.

 \medskip

 \noi For finite $q$ (including the standard non-deformed $q=1$ case),
  the time evolution is most easily described in the so called "monodromic"
 variables. Let us define them.  The phase space $M_q$ of the chiral $q$-WZW model 
 is the direct product $\lpk\times \A^1$, where $\A^1$ is the (compact)  Weyl alcove. The usual periodic parametrization of  $\lpk\times \A^1$
 is $(\ti k(\si), a^\mu H^\mu)$. In the monodromic parametrization, the points in the
 phase space $M_q$ are quasi-periodic maps $m:\br\to K$, fulfilling the monodromy condition
 $$m(\si+2\pi)=m(\si)M, \quad M=\exp{(-2\pi i a^\mu H^\mu)}.\nn$$
 The transformation from the monodromic parametrization 
 into the periodic one is given by the relations
 $$\exp{(-2\pi i a^\mu H^\mu)}=m^{-1}(\si)m(\si+2\pi), \quad \ti k(\si)=m(\si)\exp{(ia^\mu H^\mu\si)}.\nn$$
 The time evolution $E_q$ of the finite $q$ chiral WZW model is defined very simply: a point $m(\si)\in M_q$ at the time $\tau=0$
 gets evolved to the point $m(\si-\tau)$ at the time $\tau$. In the periodic parametrization,
 this time evolution gets translated into
 $$\ti k(\si)\to \ti k(\si-\tau)\exp{(ia^\mu H^\mu\tau)},\quad a^\mu \to a^\mu.\eqno(4.61)$$
 It is not difficult to see that the transformation (4.61) defines a symplectomorphism.
 This follows from the fact that the finite $q$ exchange relations (4.59) and (4.60)
 (which completely characterize the symplectic structure on $M_q$) 
 are invariant with respect to (4.61).   

 \medskip

 \noi Coming back to the  case $q\to \infty$, we  define the time  evolution in terms
 of the dual variables $(\ti k,\ti a)\in \ti M_\infty$ (cf. Eq. (4.41)).  The advantage of working with  the
 dual variables in this context is clear: they are the best adapted for the description of
 the $q\to\infty$ limit. We remind that $\ti a=e^{k\e'a^\mu H^\mu}=e^{-\phi^\mu H^\mu}$,
 which gives $a^\mu=-\frac{\phi^\mu}{\e' k}$.   Thus the $E_q$ evolution transformation (4.61) can be rewritten
 as
  $$\ti k(\si)\to \ti k(\si-\tau)\exp{(-\frac{i\phi^\mu}{\e' k} H^\mu\tau)},\quad \phi^\mu \to \phi^\mu.\nn$$
 The $\e'\to -\infty$ limit is achieved by keeping $\phi^\mu$ fixed which suggests the following
 time evolution $E_\infty$ of the chiral $\infty$-WZW model:
   $$\ti k(\si)\to \ti k(\si-\tau), \quad \phi^\mu \to \phi^\mu.\eqno(4.62)$$
  Let us  verify, that the suggested time evolution (4.62) is correctly defined, which means that 
 it leaves invariant the subspace $\ti M_\infty\subset \lpk\times A_-$ and it defines the one-parameter
 group of symplectomorphisms of $\ti M_\infty$. The latter statement follows from
 the obvious invariance of the dual exchange relations (4.57), (4.58)  with respect to  the transformation (4.62). To prove the  former statement is also easy.   Indeed, let $l(\si)$
  be an element of $\lpkc$ which is also in the domain of definition of the maps $\Lm_R$ and
  $\Xi_L $, i.e. $l(\si)\in S_\infty$. Then it follows that also $l(\si-\tau)$ is in $S_\infty=G_RG^*$, since 
  the transformation of $D=\lpkc$ defined by $l(\si)\to l(\si-\tau)$ leaves the subgroups $G_R$ and $G^*$    invariant.   

    \medskip

 \noi In this section, we have completed  the definition of the 
 chiral $\infty$-WZW model by  defining the  consistent time evolution $E_\infty$  on
 the phase space $(\ti M,\ti\Om_\infty)$.   However, we may ask another question: is this 
 evolution generated by a Hamiltonian function? The answer to this question is affirmative provided
 that we enlarge   slightly the phase space  $M_\infty$ by working with the smooth loop groups
 $LK^\bc$, $LK$ etc.
 rather than with the polynomial ones $\lpkc$, $\lpk$ etc.
 Now recall the result described in Sec 8.9 of \cite{PS}  that  every infinitesimal symplectomorphism of
an infinite-dimensional simple connected symplectic manifold has a Hamiltonian function. Let us therefore argue 
that the chiral $\infty$-WZW phase space $\ti M_\infty$ is simply connected. First of all, we know that
$\ti M_\infty$
is diffeomorphic to $M_\infty=G_R\times A_-$  which gives the following relation
between the fundamental groups:
$$\pi_1(\ti M_\infty)=\pi_1(G_R)\times \pi_1(A_-).\nn$$
Obviously,  $A_-$ is   simply connected and
it  remains to show that $G_R$ is simply connected.  

\medskip

\noi Denote respectively  by
$L^0K^\bc$ and $L^0K$ the subgroups of $LK^\bc$ and $LK$ formed of the loops
verifying $l(0)=e_{K^\bc}$ and $k(0)=e_K$.  Similarly, denote by $G_R^0$ the subgroup
of $G_R$ for which $\gamma_0=e_{K^\bc}$ (cf. the expansion (4.2)).   Clearly, $LK^\bc$ is diffeomorphic to $K^\bc \times L^0K^\bc$,
$LK$ is diffeomorphic to $K \times L^0K$ and  $G_R$ is diffeomorphic to $K \times G_R^0$. Thus we obtain
$$\pi_1(LK^\bc)=\pi_1(K^\bc)\times \pi_1(L^0K^\bc)= \pi_1(K^\bc)\times \pi_2(K^\bc),\eqno(4.63)$$
$$\pi_1(LK)=\pi_1(K)\times \pi_1(L^0K)= \pi_1(K)\times \pi_2(K),\eqno(4.64)$$
$$\pi_1(G_R)=\pi_1(K)\times \pi_1(G_R^0).\eqno(4.65)$$
Note also that the global decomposition $LK^\bc=G^*LK$ (cf. Lemma 6)  implies also
a  global decomposition $LK^\bc = G_0^* (AN) LK$, where $G_0^*$
is the subgroup
of $G^*$ for which $\gamma_0=e_{K^\bc}$.
  Taking the hermitian conjugation of the global
decomposition $LK^\bc = G_0^* (AN) LK$, we establish that $LK^\bc$ is diffeomorphic
to $LK \times (AN)^\dagger \times G_R^0$. Hence we have
$$\pi_1(LK^\bc)=\pi_1(G_R^0)\times \pi_1((AN)^\dagger))\times \pi_1(LK).\eqno(4.66)$$
We have supposed that $K$ is simple connected and simply connected, which means
that the fundamental groups $\pi_1(K)$, $\pi_1(K^\bc)$ and $\pi_1((AN)^\dagger)$ are trivial.
Moreover, the classical theorem says (cf. Sec 8.6 of \cite{PS}) that the second homotopy groups $\pi_2(K)$ and
$\pi_2(K^\bc)$ are also trivial.  Thus we deduce from the relations (4.63-66) the desired
result  that the fundamental group of $G_R$ is also trivial.

\medskip

\noi We have established the existence of the hamiltonian generating the simple time
evolution (4.62), however, we do not have any explicit formula expressing it as the
function of $\ti k$ and $\ti a$.  This  type of situation took place  also for the finite $q$ case; it was
only for the standard $q=1$ WZW model that such an explicite (i.e. Sugawara) formula
could have been written.

  \section{Conclusions and outlook}

  \medskip

 \noi The  paper is devoted to the
explicit construction of the dynamical system, which can be interpreted
as the $q\to\infty$ limit of the  $q$-WZW model.  The most important
feature of this system is its symmetry structure; in fact, 
 the limiting model has two {\it non-isomorphic} symmetry groups $G_L$ and $G_R$, which is not true for  the finite $q$ WZW model  (including the
standard case $q=1$),
thus we may even   say 
 that the symmetry structure of  the $q\to \infty$   WZW model
 is richer and more
 intriguing than that of its finite $q$ counterpart.  In order to identify the symmetries of the model,
we have used in  its full richness
the concept of the affine Poisson group and of the anomalous Poisson-Lie symmetry.
  In fact, the formulation of the  $q\to\infty$ WZW model based on the notion
   of the symplectic grupoid of the affine Poisson group was very insightful;
in particular,  we would not have suspected the existence of the
remarkable duality transformation (4.41) if we did not take
the inspiration from Lemma 5.

\medskip

\noi There are also aspects of the structure of the $q\to\infty$ WZW model
which did not follow from the general theory of the affine Poisson groups, like e.g. the chiral decomposability.  
 The possibility
to study the simpler chiral $q\to\infty$ WZW model, which also
enjoys the anomalous Poisson-Lie symmetry, has
lead to very explicit formulae like the exchange relations
(4.57)  and (4.58)  and was also very 
suggestive on general grounds.
 In particular, we found
remarkable that the $q\to\infty$ trigonometric
exchange relations (4.57) and (4.58) 
are so much simpler
than the finite $q$ elliptic exchange relations (4.59) and (4.60).

\medskip

\noi Finally, we believe that the quantization of the results
of the present paper will be doable due to  the simplicity of the trigonometric exchange relations (4.57) and
(4.58). The eventual quantization will probably make a fruitful use of
  the theory of twisting  of Hopf algebras 
developed by Majid \cite{Ma}.

\end{document}